\pdfoutput=1
\documentclass[sigconf,natbib=true,nonacm=true]{acmart}

\usepackage{multirow}
\usepackage{multicol}
\usepackage{amsmath}
\usepackage{comment}
\usepackage{soul}
\usepackage{xcolor}
\usepackage{wasysym}
\usepackage{titlesec}
\usepackage{subcaption}
\usepackage{url}
\usepackage{graphicx}
\usepackage{wrapfig}
\usepackage{paralist}
\usepackage{enumerate} 
\usepackage{enumitem}
\setlist{nolistsep}
\usepackage{colortbl}
\usepackage{sidecap}
\usepackage{adjustbox}
\usepackage{tikz}
\usepackage{pgfplots}
\usepgfplotslibrary{statistics}
\usetikzlibrary{pgfplots.groupplots}
\selectcolormodel{cmyk}
\usepackage{pgfplotstable}
\usepackage{chronology}
\usepackage{booktabs}

\pgfplotsset{every axis/.append style={
                    xlabel={$x$},          
                    ylabel={$y$},          
                    label style={font=\sffamily\scriptsize},
                    tick label style={font=\sffamily\scriptsize},
                    xticklabel style = {font=\sffamily\scriptsize},
                    title style = {font=\footnotesize\sffamily},
                    ylabel near ticks,
                    y label style={font=\sffamily\scriptsize},
                    xlabel near ticks,
                    x label style={font=\sffamily\scriptsize},
                    legend cell align={left},
                    legend style={draw=none, font=\sffamily\scriptsize},
                    },
                    legend image code/.code={
                    \draw[mark repeat=2,mark phase=2]
                        plot coordinates {
                        (0cm,0cm)
                        (0.15cm,0cm)        
                        (0.3cm,0cm)         
                        };%
                    },
                    boxplot/hide outliers/.code={
                        \def\pgfplotsplothandlerboxplot@outlier{}%
                    }
                    }
\pgfplotsset{compat=newest}  
\pgfplotsset{
    jitter/.style={
        x filter/.code={\pgfmathparse{\pgfmathresult+(rnd-.5)*#1}}
    },
    jitter/.default=0.1
}

\newenvironment{customlegend}[1][]{%
    \begingroup
    \csname pgfplots@init@cleared@structures\endcsname
    \pgfplotsset{#1}%
}{%
    \csname pgfplots@createlegend\endcsname
    \endgroup
}%
\def\addlegendimage{\csname pgfplots@addlegendimage\endcsname}

\newcommand{\eg}{\textit{e.g.}}
\newcommand{\ie}{\textit{i.e.}}

\newcommand{\hide}[1]{}

\usepackage{tikz-qtree}

\usetikzlibrary{shapes, shapes.arrows,  patterns, arrows, decorations.markings,shadows}
\usetikzlibrary{calc}
\usetikzlibrary{decorations.pathreplacing}
\usetikzlibrary{patterns, hobby, fit, positioning,  arrows.meta}
\tikzstyle{opencircle} = [circle,minimum width=10, draw, fill=black!5, inner sep=1.5]
\tikzstyle{faintopencircle} = [circle,minimum width=10, draw=black!40, fill=black!05, inner sep=1.5, text=black!40]
\tikzstyle{itxset} = [rounded corners=3pt, draw, minimum height=12, inner sep=0]
\tikzstyle{itxsetfocus} = [rounded corners=3pt, draw, ultra thick, minimum height=12, inner sep=0]
\tikzstyle{internalnode} = [circle, draw, fill, minimum size=1.5mm,inner sep=0pt,outer sep=0pt]
\tikzstyle{externalnode} = [circle, draw, minimum size=1.5mm,inner sep=0pt,outer sep=0pt]
\tikzstyle{nonterminal} = [draw, inner sep=1.5]
\tikzstyle{graphletnode} = [circle, draw, fill, minimum size=1.0mm,inner sep=0pt,outer sep=0pt]
\tikzstyle{splitnode} = [circle, draw, fill, minimum size=1.2mm,inner sep=0pt,outer sep=0pt]
\tikzstyle{faintnonterminal} = [draw=black!40, inner sep=1.5, text=black!40]

\colorlet{mycyan}{cyan!30!white}

\tikzstyle{textnode} = [rectangle, inner sep=0pt,outer sep=0,execute at begin node={\strut}, font=\small]  
\tikzstyle{bnode} = [circle, draw, fill=black, minimum size=4mm, text=white, outer sep=1.5pt]  
\tikzstyle{enode} = [circle, drop shadow, draw=black, thick, inner sep=1.5pt, minimum size=15pt, fill=gray!10, outer sep=2pt, text=black]  
\tikzstyle{blueenode} = [enode, fill=mycyan]  
\tikzstyle{pinkenode} = [enode, fill=pink]  

\tikzstyle{inode} = [circle, thick, draw, minimum size=2mm, outer sep=1.5pt]  
\tikzstyle{blueinode} = [inode, fill=mycyan]  
\tikzstyle{pinkinode} = [inode, fill=pink]  

\tikzstyle{enode-hrg} = [circle, draw, fill=black, text=white, minimum size=4mm, inner sep=0.25pt, outer sep=1.5pt]  
\tikzstyle{hyperedge} = [draw, inner xsep=0.5, fill=gray!5, diamond, minimum size=0.5mm, outer sep=0.5pt]  
\tikzstyle{nt} = [draw, inner xsep=1.5, fill=gray!5, minimum size=3mm, minimum size=1mm, outer sep=1.5pt]  
\tikzstyle{tnode} = [minimum size=5mm, font=\large]  
\tikzstyle{hnode} = [enode, very thick, draw=blue, outer sep=1.5pt]  
\tikzstyle{graphletnode} = [circle, draw, fill=gray!70, minimum size=1.5mm,inner sep=0pt,outer sep=0.25pt]
\tikzstyle{hidden} = [draw=white]
\tikzstyle{edge} = [thick]  
\tikzstyle{iedge} = [edge, very thick, draw=blue]   
\tikzstyle{bedge} = [edge, draw=red]  
\tikzstyle{faded} = [opacity=0.65, text opacity=0.65, thin,]
\tikzstyle{ledge} = [thick]  

\usepackage{mathtools}
\usepackage{wrapfig}

\usepackage[english]{babel}

\newcommand{\para}[1]{\vspace{0.1cm}\noindent\textbf{#1}\quad}
\newcommand\update[1]{#1}

\usepackage[utf8]{inputenc}
\usepgfplotslibrary{groupplots,dateplot}
\usetikzlibrary{patterns,shapes.arrows}
\pgfplotsset{compat=newest}

\definecolor{seablue}{rgb}{0.631, 0.788, 0.957}
\definecolor{seaorange}{rgb}{1.0, 0.706, 0.51}
\definecolor{seagreen}{HTML}{65D965}
\definecolor{seared}{rgb}{1, 0.125, 0.082}
\definecolor{seapurple}{HTML}{9477CB}
\definecolor{seabrown}{rgb}{0.871, 0.733, 0.608}
\definecolor{seapink}{rgb}{0.98, 0.69, 0.894}

\author{Satyaki Sikdar}
\orcid{0000-0003-1669-6594}
\email{ssikdar@nd.edu}
\affiliation{
\institution{University of Notre Dame}
\country{}
}

\author{Neil Shah}
\email{nshah@snap.com}
\affiliation{
\institution{Snap Inc.}
\country{}
}

\author{Tim Weninger}
\orcid{0000-0003-3164-2615}
\email{tweninge@nd.edu}
\affiliation{
\institution{University of Notre Dame}
\country{}
}

\newif\iffast
\fastfalse

\title{Attributed Graph Modeling with Vertex Replacement Grammars}

\begin{document}

\begin{abstract}
Recent work at the intersection of formal language theory and graph theory has explored graph grammars for graph modeling. However, existing models and formalisms can only operate on homogeneous (\ie, untyped or unattributed) graphs. We relax this restriction and introduce the Attributed Vertex Replacement Grammar (AVRG), which can be efficiently extracted from heterogeneous (\ie, typed, colored, or attributed) graphs. Unlike current state-of-the-art methods, which train enormous models over complicated deep neural architectures, the AVRG model is unsupervised and interpretable. It is based on context-free string grammars and works by encoding graph rewriting rules into a graph grammar containing graphlets and instructions on how they fit together. We show that the AVRG can encode succinct models of input graphs yet faithfully preserve their structure and assortativity properties. Experiments on large real-world datasets show that graphs generated from the AVRG model exhibit substructures and attribute configurations that match those found in the input networks.
\end{abstract}


\maketitle

\section{Introduction}

One of the principal goals of graph mining is to discover patterns found in real-world graphs and use them to make predictions about various relational systems. Many such systems consist of differently typed components; it has long been argued that models which treat graphs as homogeneous, without distinguishing different types of objects and links, ignore important information. By leveraging these rich semantics as \emph{attribute labels} on nodes and edges, researchers have developed attributed graph analysis techniques to better model such data (\eg, via heterogeneous information networks),  leading to improvements in downstream inference tasks~\citep{sun2013mining, sun2011pathsim}.   Because graphs encode interaction information about many real-world relational systems, learning how to generatively model such graphs offers a promising window into their dynamics, spurring a rich line of work \citep{robins2007introduction, leskovec2010kronecker, chung2002average, karrer2011stochastic}, inspiring our present one.  

\begin{figure}[t!]
    \centering
    \iffast
        \includegraphics[width=0.3\textwidth]{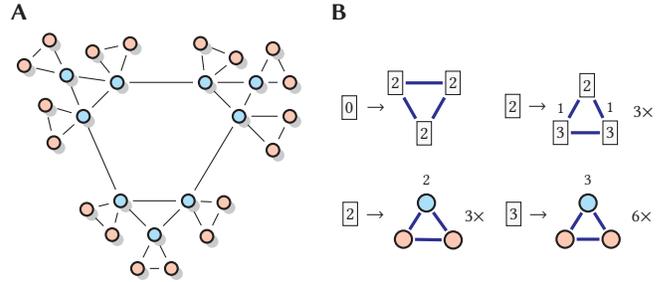}
    \else
        \begin{tikzpicture}[scale=0.9, transform shape]
\node [textnode] at (0.5,6) {\large\textbf{\textsf{A}}};
\node [textnode] at (5.2224,6) {\large\textbf{\textsf{B}}};

\begin{scope}[scale=1.25, transform shape, shift={(0,0.8)}]
\node [blueenode, minimum size=1.5mm] (1) at (1.6,1.8) {};
\node [pinkenode, minimum size=1.5mm] (2) at (1.5,1.4) {};
\node [pinkenode, minimum size=1.5mm] (3) at (1.2,1.7) {};
\node [blueenode, minimum size=1.5mm] (4) at (2,1.4) {};
\node [pinkenode, minimum size=1.5mm] (5) at (1.8,1) {};
\node [pinkenode, minimum size=1.5mm] (6) at (2.2,1) {};
\node [blueenode, minimum size=1.5mm] (7) at (2.4,1.8) {};
\node [pinkenode, minimum size=1.5mm] (8) at (2.62,1.38) {};
\node [pinkenode, minimum size=1.5mm] (9) at (2.82,1.78) {};
\node [blueenode, minimum size=1.5mm] (11) at (1.56,3.2) {};
\node [pinkenode, minimum size=1.5mm] (12) at (1.713,3.659) {};
\node [pinkenode, minimum size=1.5mm] (13) at (1.269,3.564) {};
\node [blueenode, minimum size=1.5mm] (14) at (1.16,2.8) {};
\node [pinkenode, minimum size=1.5mm] (15) at (0.811,2.488) {};
\node [pinkenode, minimum size=1.5mm] (16) at (0.711,2.931) {};
\node [blueenode, minimum size=1.5mm] (17) at (0.962,3.285) {};
\node [pinkenode, minimum size=1.5mm] (18) at (0.46,3.4) {};
\node [pinkenode, minimum size=1.5mm] (19) at (0.76,3.7) {};
\node [blueenode, minimum size=1.5mm] (21) at (3.2,3.2) {};
\node [pinkenode, minimum size=1.5mm] (22) at (3.4,3.6) {};
\node [pinkenode, minimum size=1.5mm] (23) at (3.6,3.2) {};
\node [blueenode, minimum size=1.5mm] (24) at (3,2.8) {};
\node [pinkenode, minimum size=1.5mm] (25) at (3.4,2.4) {};
\node [pinkenode, minimum size=1.5mm] (26) at (3.6,2.8) {};
\node [blueenode, minimum size=1.5mm] (27) at (2.6,3.2) {};
\node [pinkenode, minimum size=1.5mm] (28) at (2.964,3.491) {};
\node [pinkenode, minimum size=1.5mm] (29) at (2.544,3.662) {}; 


  \draw [] (1) -- (2);
  \draw [] (1) -- (3);
  \draw [] (1) -- (4);
  \draw [] (1) -- (7);
  \draw [] (1) -- (14);
  \draw [] (2) -- (3);
  \draw [] (4) -- (7);
  \draw [] (4) -- (5);
  \draw [] (4) -- (6);
  \draw [] (7) -- (8);
  \draw [] (7) -- (9);
  \draw [] (7) -- (24);
  \draw [] (14) -- (17);
  \draw [] (14) -- (15);
  \draw [] (14) -- (16);
  \draw [] (5) -- (6);
  \draw [] (8) -- (9);
  \draw [] (24) -- (27);
  \draw [] (24) -- (25);
  \draw [] (24) -- (26);
  \draw [] (11) -- (14);
  \draw [] (11) -- (12);
  \draw [] (11) -- (13);
  \draw [] (11) -- (17);
  \draw [] (11) -- (27);
  \draw [] (12) -- (13);
  \draw [] (17) -- (18);
  \draw [] (17) -- (19);
  \draw [] (27) -- (28);
  \draw [] (27) -- (29);
  \draw [] (15) -- (16);
  \draw [] (18) -- (19);
  \draw [] (21) -- (24);
  \draw [] (21) -- (27);
  \draw [] (21) -- (22);
  \draw [] (21) -- (23);
  \draw [] (22) -- (23);
  \draw [] (25) -- (26);
  \draw [] (28) -- (29);
\end{scope}

\begin{scope}[shift={(1.0116,1.94)}, scale=0.85, transform shape]
\begin{scope}[shift={(-0.0617,0.1852)}]
\node [nt] at (5.2,2.9782) {$0$};
\node at (5.6618,2.9782) {$\rightarrow$};

\node [nt] (a) at (6.5,2.5309) {$2$};
\node [nt] (b) at (7,3.4073) {$2$};
\node [nt] (c) at (6,3.4073) {$2$};
\draw [iedge] (a) -- (b) -- (c) -- (a);
\end{scope}

\begin{scope}[shift={(2.7161,1.588)}]

\node [nt] at (5.2617,1.6022) {$2$};
\node at (5.6927,1.58) {$\rightarrow$};

\node [nt] (a) at (6.5545,1.9629) {$2$};
\node [nt] (b) at (6.0927,1.1382) {$3$};
\node [nt] (c) at (6.9545,1.1382) {$3$};
\draw [iedge] (a) -- (b) -- (c) -- (a);

\node [textnode] at (6.0927,1.5513) {$1$};
\node [textnode] at (6.9545,1.5622) {$1$};
\node at (7.531,1.5) {$3\times$};
\end{scope}

\begin{scope}[shift={(-2.2766,-0.2666)}]
\node [nt] at (7.4309,1.58) {$2$};
\node at (7.8618,1.5556) {$\rightarrow$};
\node [blueinode] (v3) at (8.7545,1.7709) {};
\node [pinkinode, minimum size=1mm] (v1) at (8.3545,1.14) {};
\node [pinkinode] (v2) at (9.2191,1.1264) {};
\draw [iedge] (v1) -- (v2) -- (v3) -- (v1);
\node [textnode] at (8.7545,2.1549) {$2$};
\node at (9.5928,1.54) {$3\times$};
\end{scope}

\begin{scope}[shift={(0.5629,-0.2666)}]
\node [nt] at (7.4309,1.58) {$3$};
\node at (7.8618,1.5556) {$\rightarrow$};
\node [blueinode] (v3) at (8.7236,1.7709) {};
\node [pinkinode, minimum size=1mm] (v1) at (8.3236,1.14) {};
\node [pinkinode] (v2) at (9.1236,1.14) {};
\draw [iedge] (v1) -- (v2) -- (v3) -- (v1);
\node [textnode] at (8.7236,2.1549) {$3$};
\node at (9.6545,1.54) {$6\times$};
\end{scope}
\end{scope}

\end{tikzpicture}
    \fi
    \caption{(A) Example attributed graph and (B) an Attributed Vertex Replacement Graph Grammar (AVRG) extracted from the example graph. The AVRG succinctly and faithfully describes graph topology and attribute configuration of the input graph.}
    \label{fig:attexample}
\end{figure}

\para{Present Work.} Our work\footnote{The source code is available at \url{https://github.com/satyakisikdar/Attributed-VRG}} lies at the intersection of attributed graph mining and graph generation.  Specifically, we build upon a class of local topology-aware generators called Vertex Replacement Grammars (VRGs).   Having their origins in formal language theory, where objects of different types (\eg, characters, glyphs, words, parts of speech), graph grammars offer a promising avenue in modeling attributed graphs like the small example illustrated in~\autoref{fig:attexample}(A). 
Attributed Vertex Replacement Grammar (AVRG) builds on recent advances in graph grammars~\citep{hibshman2019towards, sikdar2019modeling,aguinaga2018learning}, which were designed only to model and generate homogeneous graphs, to handle \textit{attributed} graphs, increasing its applicability to a broad class of modern problems. 

Graph neural networks (GNNs) comprise a wide swath of models with similar goals in attributed graph modeling. These models train deep neural network architectures using loss functions and stochastic gradient descent. GNNs offer excellent modeling capacity in downstream machine learning tasks but can be difficult to train, have enormous parameter spaces, and do not readily permit human inspection. 
Despite performing well on link prediction and node classification tasks, some GNN architectures fail to preserve the topological fidelity of the input graphs~\citep{sikdar2020infinity}. AVRGs, on the other hand, can capture and model the subtle intricacies in both topological and attribute spaces without requiring supervision, deep neural architectures, and training.

Simply put, the goal of the present work is to extract graph rewriting rules, like those illustrated in \autoref{fig:attexample}(B), that are representative of the topology and attribute configuration of the original graph. Our key idea is to use an exemplar input graph to build a grammar of graph rewriting rules with patterns that can be applied stochastically to generate new, diverse graphs that meet desirable attribute and topology-related graph properties.  Both formally and through a running example, we introduce and describe the \textbf{Attributed Vertex Replacement Grammar} (AVRG) formalism and show the following:
\begin{enumerate}[topsep=2pt]
    \item AVRGs can extract interesting and frequently appearing attributed subgraphs from large real-world graphs,  
    
    \item AVRGs can generate graphs that match the topology and attribute assortativity of the original graph, and 
    
    \item the ability of AVRGs to handle a wide variety of assortative and disassortative graphs. 
\end{enumerate}

Although graph grammars appear to be a natural fit for modeling attributed graphs, this task presents many challenges. Our paper describes many of the modeling decisions made as a result and their consequences. We also demonstrate the power and flexibility of AVRGs with two case studies--the former focuses on the ability to extract meaningful patterns in~\autoref{sec:real-world-avrg}, and the latter on testing the generative ability of the model on well-known attributed graphs with various degree and attribute assortativity values in~\autoref{sec:cabam}.
We find that AVRG models provide a succinct yet faithful representation of input graphs--often better than attribute-based probabilistic models and graph neural network models at a far smaller computational cost and model size. 


\vspace{-0.2cm}
\section{Related work}

Attributed graphs are at the core of most modern graph mining literature.  Considerable prior work focuses on clustering~\citep{zhou2010clustering, perozzi2014focused}, classification~\citep{xu2018powerful}, anomaly detection~\citep{shah2016edgecentric}, similarity~\citep{su2016fast, al2019ddgk}, and search~\citep{tong2007fast} tasks on attributed graphs.  The end goal of such methods is usually \emph{descriptive} or \emph{inferential}. On the one hand, descriptive methods tend to focus on discovering, enumerating, or counting patterns in given graph data~\citep{grahne2005fast,ahmed2015efficient}, which may be attributed~\citep{yan2002gspan,sun2012efficient} or dense~\citep{hooi2016fraudar, nilforoshan2019slicendice}.  On the other hand, inferential methods involve statistical reasoning over graph data to make educated inferences (\eg, on nodes, edges, or graphs) using examples ~\citep{perozzi2014deepwalk, tang2015line, kipf2016semi,velivckovic2017graph,xu2018powerful,dong2017metapath2vec, gao2018deep,zhu2020graph,mcpherson2001birds,newman2003mixing,zhu2020beyond, zhu2020graph}. 

Other works study graphs through a \emph{prescriptive} lens by characterizing their formation and generative processes, usually from a mathematical perspective.  This subfield, called graph generation, typically involves positing generative processes that aim to produce graphs with desirable (usually topological) statistical properties that match real-world phenomena. These include exponential random graphs~\citep{robins2007introduction}, Kronecker graphs~\citep{leskovec2010kronecker}, Chung-Lu graphs~\citep{chung2002average}, and Stochastic Block Models (SBMs)~\citep{karrer2011stochastic} involves learning parameters via likelihood maximization based on a given input graph, which can be used to generate new graphs mimicking the input graph's properties.  Several very recent works also involve neural graph generation with recurrent neural networks~\citep{you2018graphrnn}, variational autoencoders~\citep{simonovsky2018graphvae}, and
generative adversarial networks~\citep{bojchevski2018netgan}.  However, most graph generation literature focuses on non-attributed graphs.  The few exceptions~\citep{pfeiffer2014attributed,pfeiffer2012fast,mussmann2014assortativity} adapt Chung and Lu's model for attributed graphs; yet, they ignore local graph topology, which is highly diverse across nodes in real graphs \citep{dong2017structural} with implications for tasks like anomaly detection~\citep{akoglu2010oddball}, social contagion and diffusion~\citep{ugander2012structural,cheng2014can}, and engagement prediction~\citep{yang2018know}.

A recent addition to the class of graph generators are hyperedge~\citep{aguinaga2016growing,aguinaga2018learning} and node replacement~\citep{hibshman2019towards,sikdar2019modeling} grammars. Graph grammars contain graphical rewriting rules that match and replace graph fragments, similar to how a context-free string grammar rewrites characters in a string. These graph fragments represent a succinct description of the network's building blocks, and the graph grammar's rewiring rules describe the instructions about how the graph is pieced together. 
Unfortunately, existing grammar formalisms do not provide for attributes, nor do they capture complex attribute mixing patterns during graph generation, making them unsuitable for attributed graphs. 



\section{Attributed Vertex Replacement Grammars}

\subsection{Preliminaries}
\para{Labeled Attributed Multigraphs.} 
A labeled attributed multigraph is a $5$-tuple $H = \langle V, E, \kappa, L, \mathcal{A} \rangle$ where $V$ is the set of vertices; $E \subseteq V \times V$ is the set of edges; $\kappa : E \mapsto \mathbb{Z}^{+}$ is a function assigning multiplicity to edges; $L$ is the set of labels on nodes; and $\mathcal{A}: V \mapsto L$ is a function mapping nodes to attribute values in $L$.
By default, each edge has a multiplicity value of 1. Although our model can be used for directed graphs, the present work treats all graphs as undirected for clarity of prose and illustration.
Attributes can be internal, based on the graph topology like node degree, or external, based on externally observed parameters, such as age or gender.
We only focus on discrete attributes for simplicity, but these models can usually be extended relatively easily to handle continuous values.
We use $|\mathcal{A}|$ to denote the number of unique attribute values, so for the graph in~\autoref{fig:attexample}(A), $|\mathcal{A}| = 2$ because it contains two different types/colors of nodes: \emph{blue} and \emph{pink}. 

\para{Assortativity.} Assortativity functions permit us to quantify the nature of mixing patterns of attributes in networks. In this work, we adopt the definitions proposed by \citeauthor{newman2003mixing} ~\citep{newman2003mixing}. High assortativity values represents homophily, the \textit{birds-of-a-feather} principle~\citep{mcpherson2001birds}, where nodes are likely to connect to nodes with similar attributes.
%
Typically graph modelers are interested in degree assortativity. That is, the assortativity of a network where node degrees were considered to be attributes. In fact, in much of the related literature, the term \textit{assortativity} is generally taken to imply degree assortativity. However, in the present work, we endeavor to be precise when referring to the domain. 
Table~\ref{tab:datasets} describes 12 datasets that are used as the experimental corpora in the present work. Each dataset is labeled with the number of nodes $|V|$, edges $|E|$, and number of unique attributes $|\mathcal{A}|$; as well as the degree assortativity $\rho_\text{deg}$ and external attribute assortativity $\rho_\text{attr}$.

\para{Normalized Dasgupta Cost (NDC).} We judge the goodness of a dendrogram $\mathcal{D}$ obtained from hierarchical graph clustering algorithms using the $(NDC)$ metric defined below~\citep{dasgupta2016cost}.
\vspace*{-7pt}
\begin{equation}
 NDC(\mathcal{D},G) = \frac{\sum_{(u, v)\, \in\, E} |\text{leaves}(\text{LCA}(u, v, \mathcal{D}))|}{|V| \cdot |E|} \label{eq:norm-dasgupta}
\end{equation} 
where $|\text{leaves}({\text{LCA}(u, v, \mathcal{D})})|$ is the number of leaves in the subtree rooted by lowest common ancestor (LCA) of nodes $u$ and $v$.
A good dendrogram $\mathcal{D}$ is expected to have a low cost because connected nodes in the graph should share an LCA lower down in the tree. 




\subsection{The AVRG Formalism}

An Attributed Vertex Replacement Grammar (AVRG) is a $4$-tuple $G = \langle \Sigma, \Delta, \mathcal{P}, \mathcal{S} \rangle$ where $\Sigma$ is the alphabet of node labels; $\Delta \subseteq \Sigma$ is the alphabet of \textit{terminal} node labels; $\mathcal{P}$ is a finite set of productions rules ($\mathcal{R}$) of the form $X \rightarrow (R, f)$, where $X$ is a left-hand side (LHS) consisting of a \textit{nonterminal} node (\ie, $X\in \Sigma \setminus \Delta$) with a size $\omega$; and the tuple $(R, f)$ represents the right-hand side (RHS) of the production rule, where $R$ is a labeled attributed multigraph with terminal and possibly nonterminal nodes, and $f \in \mathbb{Z}^+$ is the rule frequency, \ie, the number of times it appears in the grammar; and $\mathcal{S}$ is the starting graph $\fbox{0}$, a nonterminal of size $0$. 

Fig~\ref{fig:attexample}(B) shows an example AVRG with four rules. Each rule contains an LHS and RHS. Squares containing numbers are nonterminal-nodes that can be replaced by a rule's RHS. Colored nodes on the RHS subgraphs represent terminal nodes from the input graph. Every node in $R$ is labeled by the number of boundary edges adjacent to it in the original graph. The sum of boundary degrees on the RHS is defined as $\omega$ and used to label the LHS.

Incorporating attributes in the formalism introduces new challenges in both extraction and generation; we elaborate next.

\section{Extracting AVRGs}
\label{sec:extract}
At a high level, our goal in the extraction process is to identify small but meaningful building blocks of a graph while also keeping track of how they can be pieced together. One way to find such building blocks would be to exhaustively enumerate all possible subgraphs up to a certain size, but this quickly becomes infeasible for even medium-sized graphs.
We opt to use a dendrogram obtained from a hierarchical clustering algorithm to help us find the building blocks to make our model scalable for large graphs. 
This works well in practice because most real-world graphs are known to be self-similar across multiple scales, and therefore their nodes can be decomposed into clusters in a hierarchical fashion~\citep{ravasz2003hierarchical}.
Once we discover the blocks, we extract grammar rules and contract both the graph and the dendrogram. We repeat this extraction and contraction process until the graph is empty.
It is important to note that we only run the clustering algorithm once at the beginning and not every time the graph gets updated.

\subsection{Selecting a \textit{Good} Clustering Algorithm}
\label{apx:cluster}
\begin{figure}[t]
    \centering
    \iffast
        \includegraphics[width=0.3\textwidth, height=0.21\textheight]{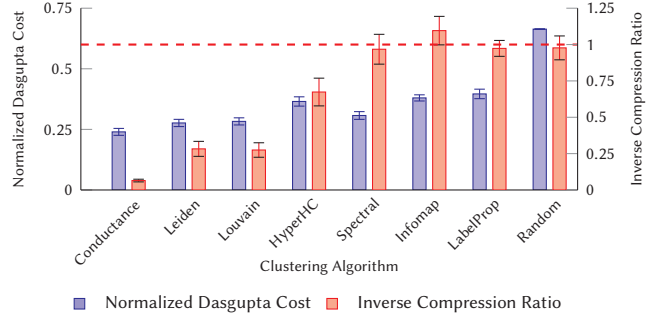}
    \else
        \pgfplotstableread{
clustering  cost    cost95 invcompression  invcompression95
Conductance 0.2397 0.0143 0.0648 0.0092
HyperHC 0.3652 0.0192 0.6736 0.0953
Infomap 0.3799 0.0127 1.0959 0.0979
LabelProp 0.3963 0.0196 0.9736 0.0547
Leiden 0.2765 0.0149 0.2827 0.0516
Louvain 0.283 0.0147 0.2748 0.0499
Random 0.6637 0.0014 0.9772 0.0821
Spectral 0.3073 0.016 0.9678 0.1026
}{\CostCompression}

\begin{tikzpicture}
    \begin{axis}[
        width  = 0.46\textwidth,
        axis y line*=left,
        height = 4cm,
        major x tick style = transparent,
        axis on top,
        point meta=explicit symbolic,
        ybar=5*\pgflinewidth,
        nodes near coords,
        bar width=5pt,
        ylabel = {\scriptsize{\textsf{Normalized Dasgupta Cost}}},
        xlabel = {\textsf{Clustering Algorithm}},
        symbolic x coords={Conductance,Leiden,Louvain,HyperHC,Spectral,Infomap,LabelProp,Random},
        xtick = data,
        xticklabel={\textsf{\tick}},
        x tick label style={font=\scriptsize, rotate=45,anchor=east, xshift=8pt, yshift=-5pt},
        yticklabel={$\mathsf{\pgfmathprintnumber{\tick}}$},
        y tick label style={font=\scriptsize},
        scaled y ticks = false,
        enlarge x limits=0.09,
        ymin=0,ymax=0.75,
        ytick={0,0.25,0.5,0.75},
        xlabel style={yshift=10pt},
        axis x line*=bottom,
    ]
        
        \addplot plot [error bars/.cd, y dir=both, y explicit,] table [x=clustering,y=cost, y error=cost95] {\CostCompression};

    \end{axis}

    \begin{axis}[
    axis y line*=right,
    axis x line=none,
    axis on top,
    xtick=\empty,
    width = 0.46\textwidth,
    height = 4cm,
    point meta=explicit symbolic,
    ybar,
    bar width=5pt,
    nodes near coords,
    symbolic x coords={Conductance,Leiden,Louvain,HyperHC,Spectral,Infomap,LabelProp,Random},
    enlarge x limits=0.09,
    ylabel = {\scriptsize{\textsf{Inverse Compression Ratio}}},
    xticklabel={\textsf{\tick}},
    yticklabel={$\mathsf{\pgfmathprintnumber{\tick}}$},
    y tick label style={font=\scriptsize},
    ytick={0,0.25,0.5,0.75,1,1.25},
    ymin=0, ymax=1.25,
    ]
    \addplot [draw=red, fill=red!40, xshift=1.5*\pgfplotbarwidth, nodes near coords] plot [error bars/.cd, y dir=both, y explicit] table [x=clustering,y=invcompression, y error=invcompression95] {\CostCompression};
    
    \coordinate (A) at (axis cs:Conductance,1);
    \coordinate (O1) at (rel axis cs:0,0);
    \coordinate (O2) at (rel axis cs:1,0);
    \draw [red, thick,sharp plot,dashed] (A -| O1) -- (A -| O2);
    \end{axis}
\end{tikzpicture}

\begin{tikzpicture}
    \begin{customlegend}[ 
    legend columns=2,
    legend style={
    draw=none,
    font=\footnotesize,
    column sep=1ex,
  },
  legend cell align={left},
  legend entries={~\textsf{Normalized Dasgupta Cost}~,~\textsf{Inverse Compression Ratio}~}
  ]
    \addlegendimage{draw=blue,fill=blue!40,mark=square*,scale=2,only marks}
    \addlegendimage{draw=red,fill=red!40,mark=square*,scale=2,only marks}
    \end{customlegend}
\end{tikzpicture}
    \fi
    
    \caption{Mean-average Normalized Dasgupta Cost and Inverse Compression Ratio grouped by clustering algorithm. Error bars represent 95\% confidence interval. Lower is better for both axes. Red dashed line at y=1 indicates no model compression. We pick Conductance, Leiden, and Louvain algorithms for further consideration.}
    \label{fig:clustering-selection}
\end{figure}

We select 7 popularly used hierarchical graph clustering algorithms: conductance based bisection (Conductance)~\citep{hagen1992new}, spectral clustering (Spectral)~\citep{ng2002spectral}, Hyperbolic Hierarchical Clustering (HyperHC)~\citep{chami2020trees}, Leiden~\citep{traag2019from}, Louvain~\citep{blondel2008fast}, Infomap~\citep{rosvall2009map}, and recursive label propagation (LabelProp)~\citep{raghavan2007near} algorithms. 
We run these algorithms on the graphs described in~\autoref{tab:datasets} and obtain dendrograms. 
For every dendrogram, we calculate its Normalized Dasgupta Cost as defined in Eq.~\ref{eq:norm-dasgupta}, group on clustering algorithm (\ie, average across datasets), and plot the mean cost and 95\% confidence interval (in blue) in~\autoref{fig:clustering-selection}. We adopted the formalism from~\citeauthor{sikdar2019modeling} to calculate the description length~\citep{sikdar2019modeling} and the inverse compression ratio, \ie, the ratio of the description length of a grammar and the input graph, such that lower is better. The mean inverse compression ratio and 95\% confidence interval are plotted (in red) in~\autoref{fig:clustering-selection} alongside the Normalized Dasgupta cost.
We find that Conductance, Louvain, and Leiden perform best on average, so we down-select these three clustering algorithms for the next steps.

\subsection{Selecting a Suitable Internal Tree Node}
\label{apx:nodeselect}

Let $T = \{\eta_1, \cdots, \eta_k\}$ be the set of internal nodes in the dendrogram $\mathcal{D}$.
We observe that the leaves of a subtree rooted at an internal node $\eta_i \in T$ correspond to a subset of vertices of the $H$, \ie, $V_{\eta_i} \subseteq V$.
For example, in~\autoref{fig:extract}(A), $V_{\eta_3}$ = $\{c, d, e\}$ and let $H_\eta$ be the subgraph induced by $V_\eta$ on $H$.
Considering each internal node is not particularly helpful, especially in large graphs, given the overlap in the subgraphs induced by a tree node and its ancestors.
Therefore, our goal is to determine a strategy that selects a particular tree node $\eta^\ast$ from $T$ that creates small but meaningful grammar rules.
So, let $s$ be a function $s: T \mapsto \mathbb{R}$ which assigns scores to individual tree nodes, and $T^\ast \subseteq T$ be the set of the \textit{lowest} scoring tree nodes, \ie, $T^\ast = \arg \min_{\eta \in T} \{s(\eta)\}$.
Now, depending on the scoring function and the state of the current dendrogram, $T^\ast$ may contain more than one \textit{optimal} tree node, in which case we need a selection policy.

The most straightforward policy would be to pick a tree node at \textit{random} from $T^\ast$. A slightly more sophisticated policy would consider scoring functions that operate based on the subtree's size. 
In this approach, we introduce a new parameter $\mu \in \mathbb{Z}^+$. Our goal is to extract rules where the RHS subgraphs have exactly (or nearly) $\mu$ nodes. This aligns with our goal of finding small but meaningful rules from the input graph.
A simple scoring function that enforces the size-based approach would be $s(\eta_i) = |V_{\eta_i}| - \mu$. 

Once we pick the best tree node $\eta^\ast$ $\in$ $T^\ast$, we proceed to extract a grammar rule $\mathcal{R}$ that we then add to grammar $G$.

\subsection{Extracting an AVRG Rule}
Let $H_{\eta^\ast} = (V_{\eta^\ast}, E_{\eta^\ast})$ be the subgraph induced by $V_{\eta^\ast}$ on $H$, where we copy over the corresponding attribute values from $V$ to $V_{\eta^\ast}$, and $E_{\eta^\ast} = \{(u, v)\,|\, u,v \in V_{\eta^\ast} \land (u, v) \in E \}$.
Let $E_{\text{cut}}^\ast = \{(u, v)\,|\, u \in V_{\eta^\ast} \land v \in V \}$ be the set of edges, called \textit{boundary edges}, that span from $H_{\eta^\ast}$ to $H$, and let $\mathbf{b} = [b_1\ b_2\ \cdots\ b_{|V_{\eta^\ast}|}]^T$ be the vector of \textit{boundary degrees} for each node $v \in V_{\eta^\ast}$. 
The boundary degree ($b_i$) of a node $v_i$ stores the number of boundary edges to which $v_i$ is adjacent, \ie, $b_i = |\{(u, v)\,|\, (u, v) \in E_{\text{cut}}^\ast \land (u = v_i \lor v = v_i)\}|$, and therefore $\sum \mathbf{b} = |E_{\text{cut}}^\ast| = \omega$. We include the boundary degree as an additional attribute for each node in $H_{\eta^\ast}$.

\begin{figure}[t]
    \centering
    \iffast
        \includegraphics[height=0.26\textheight]{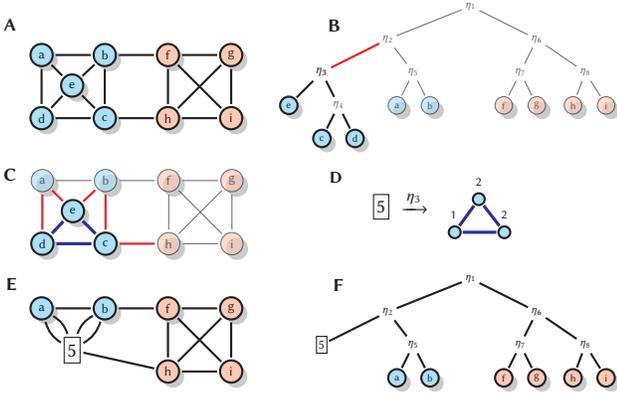}
    \else
        \begin{tikzpicture}[scale=0.8, transform shape]
\begin{scope}[node distance=0.85cm, scale=0.7, transform shape, xshift=-1.1cm, shift={(0.0397,2.8393)}]
	\node [textnode, scale=1.5] at (-2.75,1.1786) {\textsf{\textbf{A}}};
	\node [blueenode] (a) at (-2,0.5) {\textsf{a}};
	\node [blueenode] (b) at (-0.5,0.5) {\textsf{b}};
	\node [blueenode] (c) at (-0.5,-1) {\textsf{c}};
	\node [blueenode] (d) at (-2,-1) {\textsf{d}};
	\node [blueenode] (e) at (-1.275,-0.225) {\textsf{e}};
	
	\node [pinkenode] (f) at (1,0.5) {\textsf{f}};
	\node [pinkenode] (g) at (2.5,0.5) {\textsf{g}};
	\node [pinkenode] (h) at (1,-1) {\textsf{h}};
	\node [pinkenode] (i) at (2.5,-1) {\textsf{i}};
	
	\draw[edge] (a) -- (b) -- (c) -- (d) -- (a);
	\draw[edge] (a) -- (e) -- (c);
	\draw[edge] (b) -- (e) -- (d);
	\draw[edge] (c) -- (h);
	\draw[edge] (b) -- (f) -- (g) -- (h) -- (i); 
	\draw[edge] (h) -- (f) -- (i) -- (g);
\end{scope}

\begin{scope}[shift={(5.5375,1.5)}, scale=0.55, transform shape]
	\node [textnode, scale=2] at (-5.125,2.375) {\textsf{\textbf{B}}};
	\node [tnode, faded] (e1) at (-1,3) {$\eta_1$};
	\node [tnode, faded] (e2) at (-3.5,2) {$\eta_2$};
	\node [tnode, faded] (e6) at (1,2) {$\mathbf{\eta_6}$};
	\node [tnode] (e3) at (-5.5,1) {\Large $\mathbf{\eta_3}$};
	\node [tnode, faded] (e5) at (-2.7336,1) {$\eta_5$};
	\node [tnode, faded] (e7) at (0.5,1) {$\eta_7$};
	\node [tnode, faded] (e8) at (2.4672,1) {$\eta_8$};
	
	\node [blueenode, faded] (a) at (-3.2276,0) {\textsf{a}};
	\node [blueenode, faded] (b) at (-2.2276,0) {\textsf{b}};
	\node [blueenode] (c) at (-5.494,-1) {\textsf{c}};
	\node [blueenode] (d) at (-4.494,-1) {\textsf{d}};
	\node [blueenode] (e) at (-6.494,0) {\textsf{e}};
	
	\node [pinkenode, faded] (f) at (0,0) {\textsf{f}};
	\node [pinkenode, faded] (g) at (1,0) {\textsf{g}};
	\node [pinkenode, faded] (h) at (2.0922,0) {\textsf{h}};
	\node [pinkenode, faded] (i) at (3.0922,0) {\textsf{i}};	
	\node [tnode, faded] (e4) at (-4.994,0) {$\eta_4$};

	\draw [edge, faded] (e1) -- (e2);
	\draw [bedge] (e2) -- (e3);
	\draw [edge] (e3)	-- (e);
	\draw [edge] (e3) -- (e4) -- (c);
	\draw [edge] (e4) -- (d);
	\draw [edge, faded] (e2) -- (e5) -- (a);
	\draw [edge, faded] (e5) -- (b);
	
	\draw [edge, faded] (e1) -- (e6);
	\draw [edge, faded] (e6) -- (e7);
	\draw [edge, faded] (e7) -- (f);
	\draw [edge, faded] (e7) -- (g);
	\draw [edge, faded] (e6) -- (e8) -- (h);
	\draw [edge, faded] (e8) -- (i);
\end{scope}


\begin{scope}[node distance=0.85cm, scale=0.7, transform shape, xshift=-1.1cm, shift={(0.0576,-0.1434)}]
\node [scale=1.4] (subc) at (-2.7679,0.6433) {\textsf{\textbf{C}}};

	\node [blueenode, faded] (a) at (-2,0.5) {\textsf{a}};
	\node [blueenode, faded] (b) at (-0.5,0.5) {\textsf{b}};
	\node [blueenode] (c) at (-0.5,-1) {\textsf{c}};
	\node [blueenode] (d) at (-2,-1) {\textsf{d}};
	\node [blueenode] (e) at (-1.275,-0.225) {\textsf{e}};
	
	\node [pinkenode, faded] (f) at (1,0.5) {\textsf{f}};
	\node [pinkenode, faded] (g) at (2.5,0.5) {\textsf{g}};
	\node [pinkenode, faded] (h) at (1,-1) {\textsf{h}};
	\node [pinkenode, faded] (i) at (2.5,-1) {\textsf{i}};
	
	\draw[edge, faded] (a) -- (b);
	\draw[bedge] (b) -- (c);
	\draw[iedge] (c)	-- (d);
	\draw[bedge] (d) -- (a);
	\draw[bedge] (a) -- (e);
	\draw[iedge] (e) -- (c);
	\draw[bedge] (b) -- (e);
	\draw[iedge] (e) -- (d);
	\draw[bedge] (c) -- (h);
	\draw[edge, faded] (b) -- (f);
	\draw [edge, faded] (f)	-- (g) -- (h) -- (i); 
	\draw[edge, faded] (h) -- (f) -- (i) -- (g);
\end{scope}

 \begin{scope}[shift={(4.8125,2.75)}]
	\node [textnode] at (-2.0625,-2.4687) {\textsf{\textbf{D}}};
	\node [nt] at (-1.3125,-2.9369) {$5$};
	\node at (-0.75,-2.8744) {$\xrightarrow{\eta_3}$};
	
	\begin{scope}[scale=0.55, transform shape, shift={(-1.1631,-5.1097)}]
		\node [blueinode, scale=1.2] (e) at (1.7272,0) {};
		\node [blueinode, scale=1.2] (d) at (1,-1) {};
		\node [blueinode, scale=1.2] (c) at (2.5,-1) {};
		
		\draw [iedge] (c) -- (d) -- (e) -- (c);
		
		\node [textnode, scale=1.35] at (0.9544,-0.5) {\textsf{1}};
		\node [textnode, scale=1.35] at (1.7272,0.5) {\textsf{2}};
		\node [textnode, scale=1.35] at (2.5,-0.5) {\textsf{2}};
	\end{scope}
\end{scope}


\begin{scope}[node distance=0.85cm, scale=0.7, transform shape, xshift=-1.1cm, shift={(0.5397,-3.1796)}]
\node [scale=1.4] (subc) at (-3.2144,1.0895) {\textsf{\textbf{E}}};
	\node [blueenode] (a) at (-2.5,0.5) {\textsf{a}};
	\node [blueenode] (b) at (-1,0.5) {\textsf{b}};
	\node [nt, scale=1.5] (nt1) at (-1.775,-0.4929) {\textsf{5}};
	
	\node [pinkenode] (f) at (0.5,0.5) {\textsf{f}};
	\node [pinkenode] (g) at (2,0.5) {\textsf{g}};
	\node [pinkenode] (h) at (0.5,-1) {\textsf{h}};
	\node [pinkenode] (i) at (2,-1) {\textsf{i}};
	
	\draw[edge] (a) edge[bend left=20] (nt1);
	\draw[edge] (a) edge[bend right=20] (nt1);
	\draw[edge] (b) edge[bend left=20] (nt1);
	\draw[edge] (b) edge[bend right=20] (nt1);
	
	\draw [edge] (nt1) -- (h);
	
	\draw[edge] (a) -- (b);
	\draw[edge] (b) -- (f) -- (g) -- (h) -- (i); 
	\draw[edge] (h) -- (f) -- (i) -- (g);
\end{scope}

\begin{scope}[shift={(5.5375,-3.0318)}, scale=0.55, transform shape]
	\node [textnode, scale=2] at (-5,2.7728) {\textsf{\textbf{F}}};
	\node [tnode] (e1) at (-1,3) {$\eta_1$};
	\node [tnode] (e2) at (-3.5,2) {$\eta_2$};
	\node [tnode] (e6) at (1,2) {$\mathbf{\eta_6}$};
	\node [nt, scale=1.25] (e3) at (-5.5,1) {\textsf{5}};
	\node [tnode] (e5) at (-2.7336,1) {$\eta_5$};
	\node [tnode] (e7) at (0.5,1) {$\eta_7$};
	\node [tnode] (e8) at (2.4672,1) {$\eta_8$};
	
	\node [blueenode] (a) at (-3.2276,0) {\textsf{a}};
	\node [blueenode] (b) at (-2.2276,0) {\textsf{b}};
	
	\node [pinkenode] (f) at (0,0) {\textsf{f}};
	\node [pinkenode] (g) at (1,0) {\textsf{g}};
	\node [pinkenode] (h) at (2.0922,0) {\textsf{h}};
	\node [pinkenode] (i) at (3.0922,0) {\textsf{i}};	

	\draw [edge] (e1) -- (e2);
	\draw [edge] (e2) -- (e3);
	\draw [edge] (e2) -- (e5) -- (a);
	\draw [edge] (e5) -- (b);
	
	\draw [edge] (e1) -- (e6);
	\draw [edge] (e6) -- (e7);
	\draw [edge] (e7) -- (f);
	\draw [edge] (e7) -- (g);
	\draw [edge] (e6) -- (e8) -- (h);
	\draw [edge] (e8) -- (i);
\end{scope}

\end{tikzpicture}
    \fi
    \caption{One iteration of the rule extraction and graph compression process. (A) Graph $H$, (B) $\eta^\ast$ = $\eta_3$ in  $\mathcal{D}$, (C) $V_{\eta_3}$ = \{c, d, e\}, boundary edges are drawn in red, (D) Extracted grammar rule, (E) New, reduced graph $H^\prime$ with the new nonterminal $X$ of size 5, and (F) Updated dendrogram $\mathcal{D}^\prime$.}
    \label{fig:extract}
\end{figure}

\begin{figure}[tb]
    \centering
    \iffast
        \includegraphics[height=0.15\textheight]{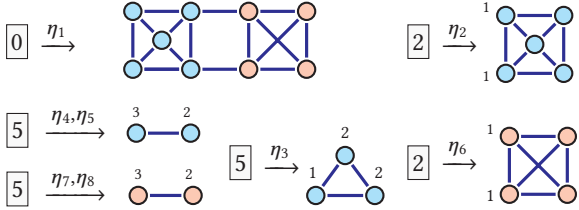}
    \else
        \begin{tikzpicture}[scale=1.1, transform shape]

\begin{scope}[shift={(0,-0.1816)}]
\node [nt] at (-2,-6.2) {$0$};
\node at (-1.5,-6.15) {$\xrightarrow{\eta_1}$};
\begin{scope}[scale=0.7, transform shape, shift={(0.1418,-2.6335)}]
	\node [blueinode] (v1) at (-1,-5.7) {};
	\node [blueinode] (v2) at (0,-5.7) {};
	\node [blueinode] (v3) at (0,-6.7) {};
	\node [blueinode] (v4) at (-1,-6.7) {};
	\node [blueinode] (v5) at (-0.5,-6.2) {};
	\node [pinkinode] (v6) at (1,-5.7) {};
	\node [pinkinode] (v7) at (2,-5.7) {};
	\node [pinkinode] (v8) at (2,-6.7) {};
	\node [pinkinode] (v9) at (1,-6.7) {};
	
	\draw [iedge] (v1) edge (v2);
	\draw [iedge] (v2) edge (v3);
	\draw [iedge](v3) -- (v4) -- (v1) -- (v5);
	\draw [iedge](v2) -- (v5) -- (v4) -- (v3) -- (v5);
	\draw [iedge](v8) -- (v9) -- (v6) -- (v8) -- (v7) -- (v9);
	\draw [iedge] (v7) edge (v6);
	\draw [iedge] (v2) edge (v6);
	\draw [iedge] (v3) edge (v9);
\end{scope}
\end{scope}

\begin{scope}[shift={(-0.95,0.4)}]
\node [nt] at (3.8,-8.2455) {$2$};
\node at (4.3,-8.15) {$\xrightarrow{\eta_6}$};

\begin{scope}[scale=0.7, transform shape, shift={(2,-3.625)}]
	\node [pinkinode] (v6) at (6,-7.7) {};
	\node [pinkinode] (v7) at (5,-7.7) {};
	\node [pinkinode] (v8) at (5,-8.7) {};
	\node [pinkinode] (v9) at (6,-8.7) {};
	\node [textnode] at (4.6714,-7.5911) {$1$};
	\node at (4.6714,-8.719) {$1$};
	
	\draw [iedge](v8) -- (v9) -- (v6) -- (v8) -- (v7) -- (v9);
	\draw[iedge]  (v7) edge (v6);
\end{scope}
\end{scope}

\begin{scope}[shift={(-5.25,-1.2545)}]
\node [nt] at (3.2545,-6.2) {$5$};
\node at (3.9545,-6.12) {$\xrightarrow{\eta_4, \eta_5}$};
	
\begin{scope}[scale=0.7, transform shape, shift={(1.9017,-2.7053)}]
	\node [blueinode] (v1) at (5.7375,-6.2) {};
	\node [blueinode] (v4) at (4.8,-6.2) {};
	\node [textnode] at (5.675,-5.85) {$2$};
	\node [textnode] at (4.8,-5.85) {$3$};
	
	\draw [iedge] (v4) edge (v1);
\end{scope}
\end{scope}

\begin{scope}[shift={(-5.2,-1.9635)}]
\node [nt] at (3.2045,-6.2) {$5$};
\node at (3.9045,-6.17) {$\xrightarrow{\eta_7,\eta_8}$};

\begin{scope}[scale=0.7, transform shape, shift={(1.964,-2.7686)}]
	\node [pinkinode] (v1) at (5.6464,-6.2) {};
	\node [pinkinode] (v4) at (4.7,-6.2) {};
	\node [textnode] at (5.575,-5.85) {$2$};
	\node [textnode] at (4.7,-5.85) {$3$};
	
	\draw [iedge] (v4) edge (v1);
\end{scope}
\end{scope}

\begin{scope}[shift={(4.65,1.8184)}]
\node [nt] at (-1.8,-8.2) {$2$};
\node at (-1.3,-8.15) {$\xrightarrow{\eta_2}$};
\begin{scope}[scale=0.7, transform shape, shift={(-0.0625,-3.5619)}]
	\node [blueinode] (v1) at (-1,-7.7) {};
	\node [blueinode] (v2) at (0,-7.7) {};
	\node [blueinode] (v3) at (0,-8.7) {};
	\node [blueinode] (v4) at (-1,-8.7) {};
	\node [blueinode] (v5) at (-0.5,-8.2) {};
	\node [textnode] at (-1.3375,-8.7589) {$1$};
	\node[textnode] at (-1.3375,-7.5661) {$1$};
	
	\draw [iedge](v1) -- (v2) -- (v3) -- (v4) -- (v1) -- (v5) -- (v2);
	\draw [iedge] (v5) edge (v4);
	\draw [iedge] (v3) edge (v5);
\end{scope}
\end{scope}

\begin{scope}[shift={(-0.3,0.4)}]
\node [nt] at (1,-8.2455) {$5$};
\node at (1.5,-8.2) {$\xrightarrow{\eta_3}$};
\begin{scope}[scale=0.7, transform shape, shift={(0.2791,-3.5619)}]
	\node [blueinode] (v3) at (3.4695,-8.7839) {};
	\node [blueinode] (v4) at (2.45,-8.7839) {};
	\node [blueinode] (v5) at (2.9565,-8.0935) {};
	
	\node at (2.4,-8.3839) {$1$};
	\node at (3.5344,-8.3839) {$2$};
	\node at (2.9565,-7.6935) {$2$};
	
	\draw [iedge](v4) -- (v3) -- (v5) -- (v4);
\end{scope}
\end{scope}
\end{tikzpicture}
    \fi
    \caption{Set of all grammar rules obtained from the dendrogram $\mathcal{D}$ for graph $H$ from~\autoref{fig:extract}(A). Note some tree nodes like $\eta_4$ and $\eta_5$ produce the same rule.}
    \label{fig:all-rules}
\end{figure}

We now have all the ingredients needed to create a grammar rule. Let $\mathcal{R} = X \rightarrow (R, f)$ be an AVRG rule, where $X$ is a nonterminal of size $\omega$, $R$ is the graph $H_{\eta^\ast}$, and the frequency $f$ = 1.
Now, if $\mathcal{R}$ is a new rule, we add it to the grammar $G$; otherwise, we update the frequency of the existing rule in $G$ that is \textit{isomorphic} to $\mathcal{R}$.
For example, we obtain 6 unique grammar rules from the dendrogram $\mathcal{D}$ in~\autoref{fig:extract}(B) which are shown in~\autoref{fig:all-rules}.

\para{Rule Isomorphism.} Two grammar rules $\mathcal{R}_1$ and $\mathcal{R}_2$ are isomorphic if and only if (a) their left-hand side nonterminals are of the same size, \ie, $\omega_1 = \omega_2$, and (b) the RHS attributed graphs are isomorphic. We use the VF2 algorithm for this calculation~\citep{cordella2004sub}. 


\subsection{Updating the Data Structures}
After extracting the grammar rule $\mathcal{R}$, we update both the current graph $H$ and dendrogram $\mathcal{D}$ to prevent the same set of nodes from directly participating in future rules.

\para{Updating the Graph.} In the current graph $H$, first we remove all the nodes in $V_{\eta^\ast}$ and its corresponding edges. Then, we introduce a new nonterminal node $X^\ast$  labeled with $\omega$ (from the rule extraction step).  
We connect $X^\ast$ to the rest of the graph through the set of cut edges $E^{\ast}_{\text{cut}}$ in $R$ where the endpoints in $V_{\eta^\ast}$ are redirected to $X^\ast$.  
Note, this may lead to the creation of multi-edges in the new graph $H^\prime$, which is now strictly smaller than $H$.  

\para{Updating the Dendrogram.} 
With a  new (smaller) $H^\prime$,  it may be prudent to re-run the clustering algorithm and generate a  new dendrogram.   However, our initial experiments found that re-clustering is time-consuming and rarely results in significant changes to the dendrogram.   Instead,  we modify $\mathcal{D}$ by replacing the subtree rooted at $\eta^\ast$ with nonterminal node $X^\ast$ labeled with $\omega$. Finally,  we set $H \leftarrow H^\prime$ and repeat the tree node selection and rule extraction processes until the dendrogram is empty.

\begin{figure}[tb]
    \centering
    \iffast
        \includegraphics[height=0.10\textheight]{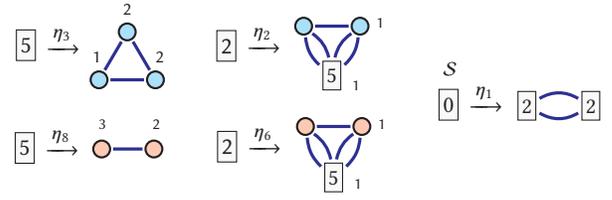}
    \else
        \begin{tikzpicture}[scale=1, transform shape]

\begin{scope}[shift={(2.976,-3.0335)}]
\node [nt] at (-5.5,5) {$0$};
\node at (-5,5.0769) {$\xrightarrow{\eta_1}$};
\begin{scope}[scale=0.75, transform shape, shift={(-1.4446,1.6482)}]
\node [nt, scale=1.25] (v1) at (-4.5,5) {$2$};
\node [nt, scale=1.25] (v2) at (-3.3666,5) {$2$};
\draw [iedge] (v2) edge[bend left] (v1);
\draw [iedge] (v2) edge[bend right] (v1);
\node [textnode, scale=1.2] at (-5.8666,5.6334) {$\mathcal{S}$};
\end{scope}
\end{scope}
\begin{scope}[shift={(-5.6578,-3.6086)}]
	\node [nt] at (-2.5,5) {\textsf{5}};
	\node at (-2,5.0769) {$\xrightarrow{\eta_8}$};
	\begin{scope}[scale=0.75, transform shape, shift={(-0.5854,1.6482)}]
		\node [pinkinode] (h) at (-0.4586,5) {};
		\node [pinkinode] (i) at (-1.3845,5) {};
		\node [textnode] at (-1.3731,5.4231) {\textsf{3}};
		\node [textnode] at (-0.4203,5.4231) {\textsf{2}};
		\draw [iedge] (h) -- (i);
	\end{scope}
\end{scope}

\begin{scope}[shift={(-6.478,-1.9724)}]
	\node [nt] at (1,4.722) {\textsf{2}};
	\node at (1.5,4.7989) {$\xrightarrow{\eta_2}$};
	
	\begin{scope}[scale=0.75, transform shape, shift={(0.7036,1.1482)}]
	\node [blueinode] (a) at (2,5.4972) {};
	\node [blueinode] (b) at (3,5.4972) {};
	\node [nt, scale=1.25] (c) at (2.5,4.6213) {\textsf{5}};
	\node[textnode] at (3.35,5.5472) {\textsf{1}};
	\node[textnode] at (2.9231,4.4231) {\textsf{1}};
	
	\draw [iedge] (a) -- (b);
	\draw [iedge] (a) edge[bend left=20] (c);
	\draw [iedge] (a) edge[bend right=20] (c);
	\draw [iedge] (b) edge[bend right=20] (c);
	\draw [iedge] (b) edge[bend left=20] (c);
	\end{scope}
\end{scope}

\begin{scope}[shift={(-6.468,-3.5918)}]
	\node [nt] at (1,5) {\textsf{2}};
	\node at (1.5,5.0769) {$\xrightarrow{\eta_6}$};
	
	\begin{scope}[scale=0.75, transform shape, shift={(0.7223,1.5741)}]
	\node [pinkinode] (a) at (2,5.451) {};
	\node [pinkinode] (b) at (3,5.451) {};
	\node [nt, scale=1.25] (c) at (2.5,4.5472) {\textsf{5}};
	\node[textnode] at (3.35,5.4731) {\textsf{1}};
	\node[textnode] at (2.9231,4.4231) {\textsf{1}};
	
	\draw [iedge] (a) -- (b);
	\draw [iedge] (a) edge[bend left=20] (c);
	\draw [iedge] (a) edge[bend right=20] (c);
	\draw [iedge] (b) edge[bend right=20] (c);
	\draw [iedge] (b) edge[bend left=20] (c);
	\end{scope}
\end{scope}

\begin{scope}[shift={(-12.1426,-2.7417)}]
\node [nt] at (4,5.5) {\textsf{5}};
\node at (4.5,5.5556) {$\xrightarrow{\eta_3}$};
\begin{scope}[scale=0.75, transform shape, shift={(1.6295,2.2964)}]
\node [blueinode] (v3) at (6,4.4231) {};
\node [blueinode] (v4) at (5,4.4231) {};
\node [blueinode] (v5) at (5.5,5.2731) {};
\node at (4.95,4.8231) {\textsf{1}};
\node at (6.0769,4.8231) {\textsf{2}};
\node at (5.5,5.6731) {\textsf{2}};
\draw [iedge] (v3) -- (v4) -- (v5) -- (v3);
\end{scope}
\end{scope}



\end{tikzpicture}
    \fi
    \caption{Grammar Rules for our example graph $H$ using the dendrogram $\mathcal{D}$ in~\autoref{fig:extract}(B). The tree nodes $\eta_3$, $\eta_2$, $\eta_8$, $\eta_6$, and $\eta_1$ where selected in order to obtain these rules. Note that $\omega$ on the LHS is equivalent to the summation of the number of broken edges on the RHS, illustrated as numbers above or to the right of each RHS-node; $f=1$ for all rules.}
    \label{fig:good-rules}
\end{figure}

In~\autoref{fig:extract} we continue the running example and illustrate the grammar rule extraction and graph and dendrogram contraction processes. The input graph (A) has 9 nodes, 16 edges, and each node is colored blue or pink. A dendrogram drawn from clustering of the graph is drawn in (B) from which $\eta_3$ is selected by using the size-based scoring function described in~\autoref{apx:nodeselect} with $\mu$=3. The subgraph representing the subtree rooted at $\eta_3$ is highlighted in (C) and contains 5 boundary edges (drawn in red). (D) illustrates a new grammar rule constructed with size-five LHS and an RHS copied from the highlighted subgraph under $\eta_3$. Finally, the extracted subgraph and selected subtree are contracted and replaced by the nonterminal node from the LHS of the newly created rule in (E) and (F). This process then repeats until the dendrogram is empty, yielding us the set of rules shown in~\autoref{fig:good-rules}.

Before discussing generation, we examine AVRGs extracted from two real graphs with distinct attribute mixing patterns. 
\subsection{Case Study I: Examining AVRG Rules} \label{sec:real-world-avrg}

\begin{figure}[tb!]
    \centering
    \pgfplotstableread{
x	y	C
		0	0	3.7
		1	0	0.02
		2	0	0.44
		3	0	0.16
		4	0	0.02
		5	0	0.79
		6	0	0.06
		0	1	0.02
		1	1	16
		2	1	0.26
		3	1	0.52
		4	1	0.61
		5	1	0.23
		6	1	0.02
		0	2	0.44
		1	2	0.26
		2	2	8.1
		3	2	0.53
		4	2	0.28
		5	2	0.74
		6	2	0.19
		0	3	0.16
		1	3	0.52
		2	3	0.53
		3	3	22
		4	3	0.66
		5	3	1.6
		6	3	1.3
		0	4	0.02
		1	4	0.61
		2	4	0.28
		3	4	0.66
		4	4	8
		5	4	0.32
		6	4	0.19
		0	5	0.79
		1	5	0.23
		2	5	0.74
		3	5	1.6
		4	5	0.32
		5	5	10
		6	5	0.87
		0	6	0.06
		1	6	0.02
		2	6	0.19
		3	6	1.3
		4	6	0.19
		5	6	0.87
		6	6	12
}{\coraheatmap}

\begin{tikzpicture}[scale=1, transform shape]
\node [textnode] at (-1,5.5344) {\LARGE \textsf{\textbf{A}}};
\node [textnode] at (-1,4.2487) {\LARGE \textsf{\textbf{B}}};
\node [textnode] at (5.7,4.287) {\LARGE \textsf{\textbf{C}}};

\begin{scope}[shift={(-0.1451,-0.0558)}, scale=0.75, transform shape]
	\begin{axis}[
	    enlargelimits=false,
	    point meta min=0,
	    every node near coord/.append style={draw=none, scale=1,
	    	text=black, yshift=-0.25cm,  /pgf/number format/.cd,
              fixed, precision=3,},
	    ymin=-0.5,ymax=6.5,
	    xmin=-0.5, xmax=6.5,
	    xtick=\empty,
	    ytick=\empty,
	    xticklabels={},
	    xlabel=,
	    ylabel=,
	    yticklabels={},
	]
        \addplot[
            matrix plot,
            fill opacity=0.5,
	     draw=white, thick,
            mesh/cols=7,
            point meta=explicit,] 
            table[x=x, y=y, meta=C] {\coraheatmap};
            
      \addplot[
            scatter,
            only marks,
            mark=none,
            nodes near coords,
            point meta=explicit,] 
            table[x=x, y=y, meta=C] {\coraheatmap};
	\end{axis}

	\node[fill=seablue, inode] (00) at (-0.45,5.25) {};
	\node[fill=seaorange, inode] (01) at (-0.45,4.475) {};
	\node[fill=seagreen, inode] (02) at (-0.45,3.65) {};
	\node[fill=seared, inode] (02) at (-0.425,2.775) {};
	\node[fill=seapurple, inode] (02) at (-0.425,2.025) {};
	\node[fill=seabrown, inode] (02) at (-0.425,1.2) {};
	\node[fill=seapink, inode] (02) at (-0.425,0.4) {};
	
	\node[fill=seablue, inode] (10) at (0.5,-0.425) {};
	\node[fill=seaorange, inode] (01) at (1.5,-0.425) {};
	\node[fill=seagreen, inode] (02) at (2.475,-0.425) {};
	\node[fill=seared, inode] (02) at (3.425,-0.425) {};
	\node[fill=seapurple, inode] (02) at (4.425,-0.425) {};
	\node[fill=seabrown, inode] (02) at (5.4,-0.425) {};
	\node[fill=seapink, inode] (02) at (6.375,-0.425) {};
\end{scope}

\begin{scope}[shift={(0.1552,5.8409)}, scale=0.8,]
	\node[fill=seablue, inode] (10) at (0.5,-0.425) {};
	\node[textnode] at (0.5,-0.9) {\textsf{5.2\%}};
	
	\node[fill=seaorange, inode] (01) at (1.75,-0.425) {};
	\node[textnode] at (1.75,-0.9) {\textsf{17.9\%}};
	
	\node[fill=seagreen, inode] (02) at (2.975,-0.425) {};
	\node[textnode] at (2.975,-0.9) {\textsf{10.5\%}};
	
	\node[fill=seared, inode] (02) at (4.2375,-0.425) {};
	\node[textnode] at (4.2375,-0.9) {\textsf{26.7\%}};
	
	\node[fill=seapurple, inode] (02) at (5.4875,-0.425) {};
	\node[textnode] at (5.4875,-0.9) {\textsf{10.1\%}};
	
	\node[fill=seabrown, inode] (02) at (6.7125,-0.425) {};
	\node[textnode] at (6.7125,-0.9) {\textsf{15\%}};
	
	\node[fill=seapink, inode] (02) at (7.9375,-0.425) {};
	\node[textnode] at (7.9375,-0.9) {\textsf{14.7\%}};
	
\end{scope}

\begin{scope}[shift={(6.1552,6.1491)}, scale=0.8]
	\begin{scope}[shift={(-9.75,-6.875)}] 
		\node[inode, fill=seared] (v1) at (9.25,3.75) {};
		\node[inode, fill=seared] (v3) at (10.75,3.75) {};
		\node[inode, fill=seared] (v2) at (10,3.75) {};
		\draw[iedge]  (v1) edge (v2);
		\draw[iedge]  (v2) edge (v3);
		\node [textnode] at (11.3685,3.75) {\textsf{24x}};
	\end{scope}
	
	\begin{scope}[shift={(-9.7203,-7.5)}]  
		\node[inode, fill=seared] (v1) at (9.65,3.75) {};
		\node[inode, fill=seared] (v2) at (10.4,3.75) {};
		\draw[iedge]  (v1) edge (v2);
		\node [textnode] at (11.3422,3.75) {\textsf{22x}};
	\end{scope}

	\begin{scope}[shift={(-9.7169,-8.275)}]  
		\node[inode, fill=seared] (v1) at (9.25,3.8) {};
		\node[inode, fill=seared] (v3) at (10.75,3.8) {};
		\node[inode, fill=seared] (v4) at (10.35,3.3) {};
		\node[inode, fill=seared] (v2) at (10,3.8) {};
		\draw[iedge]  (v1) edge (v2);
		\draw[iedge]  (v2) edge (v3);
		\draw[iedge] (v2) edge (v4);
		\draw[iedge] (v3) edge (v4);
		\node [textnode] at (11.3159,3.5) {\textsf{13x}};
	\end{scope}

	\begin{scope}[shift={(-9.75,-9.4625)}]  
		\node[inode, fill=seaorange] (v1) at (9.25,3.75) {};
		\node[inode, fill=seaorange] (v3) at (10.75,3.75) {};
		\node[inode, fill=seaorange] (v2) at (10,3.75) {};
		\draw[iedge]  (v1) edge (v2);
		\draw[iedge]  (v2) edge (v3);
		\node [textnode] at (11.3685,3.75) {\textsf{12x}};
	\end{scope}
	
	\begin{scope}[shift={(-9.7203,-10.3375)}]  
		\node[inode, fill=seapink] (v1) at (9.25,3.75) {};
		\node[inode, fill=seapink] (v3) at (10.75,3.75) {};
		\node[inode, fill=seapink] (v2) at (10,3.75) {};
		\draw[iedge]  (v1) edge (v2);
		\draw[iedge]  (v2) edge (v3);
		\node [textnode] at (11.3422,3.75) {\textsf{11x}};
	\end{scope}
	
	\begin{scope}[shift={(-9.7169,-11.0875)}]  
		\node[inode, fill=seaorange] (v1) at (9.725,3.75) {};
		\node[inode, fill=seaorange] (v2) at (10.475,3.75) {};
		\draw[iedge]  (v1) edge (v2);
		\node [textnode] at (11.3159,3.75) {\textsf{11x}};
	\end{scope}
\end{scope}

\end{tikzpicture}
    \caption{Cora. (A) Proportion of nodes in each class (color). (B) Attribute mixing matrix, each cell represents the percentage of edges with the corresponding colored endpoints. High numbers in the main diagonal and lack of off-diagonal edges indicate strong homophily. (C) The six most frequent rule RHS subgraphs of the extracted AVRG.}
    \label{fig:cora-rules}
\end{figure}
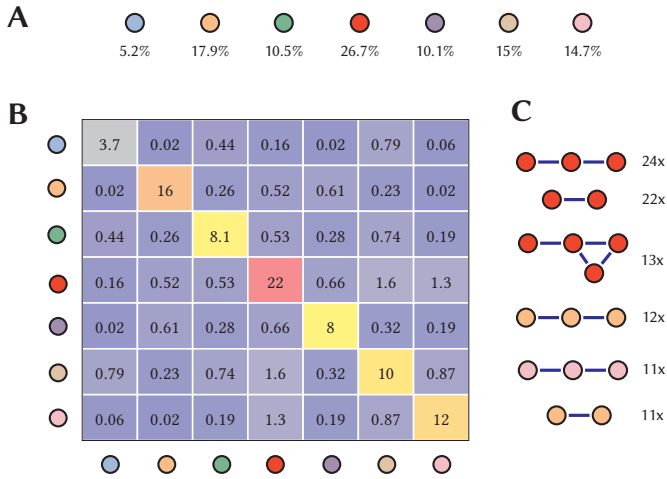

\para{Cora.}
The citation network consists of about $2500$ papers spread across $7$ different fields of Machine Learning, which are represented by colors on the nodes.
In~\autoref{fig:cora-rules}(A), we observe that not all classes are represented equally; for example, $27\%$ nodes are red while only $5\%$ nodes are blue.
The edge mixing patterns based on the node attributes in~\autoref{fig:cora-rules}(B) reflect strong homophily, with most edges spanning between papers (nodes) in the same field (color), as evidenced by high values on the main diagonal as well as the high value of $\rho_{\text{attr}} = 0.764$.
We expect the RHS subgraphs of the grammar rules extracted from Cora to capture this phenomenon. 
Indeed, in~\autoref{fig:cora-rules}(C), we find that each of the $6$ most frequent subgraphs consists of nodes belonging to the same class (color). Moreover, the top three rules all involve the red nodes, which is expected since $22\%$ of the edges in the network span between red nodes.

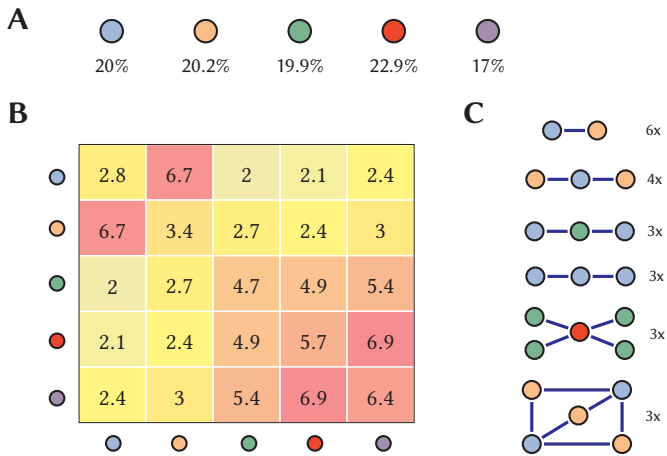
\begin{figure}[tb!]
    \centering
    \pgfplotstableread{
x	y	C
0	0	2.8
1	0	6.7
2	0	2
3	0	2.1
4	0	2.4
0	1	6.7
1	1	3.4
2	1	2.7
3	1	2.4
4	1	3
0	2	2
1	2	2.7
2	2	4.7
3	2	4.9
4	2	5.4
0	3	2.1
1	3	2.4
2	3	4.9
3	3	5.7
4	3	6.9
0	4	2.4
1	4	3
2	4	5.4
3	4	6.9
4	4	6.4
}{\chamheatmap}

\begin{tikzpicture}[scale=1, transform shape]
\node [textnode] at (-0.8159,4.8191) {\LARGE \textsf{\textbf{A}}};
\node [textnode] at (-0.8159,3.5701) {\LARGE \textsf{\textbf{B}}};
\node [textnode] at (5.25,3.5701) {\LARGE \textsf{\textbf{C}}};

\begin{scope}[shift={(-0.0023,-0.5058)}, scale=0.65]
	\begin{axis}[
	    enlargelimits=false,
	    point meta min=0,
	    ymin=-0.5,ymax=4.5,
	    xmin=-0.5, xmax=4.5,
	    xtick=\empty,
	    ytick=\empty,
	    xlabel=\empty,
	    ylabel=\empty,
	    every node near coord/.append style={
	    draw=none, scale=1.5,
	    	text=black, yshift=-0.25cm,  /pgf/number format/.cd,
              fixed, precision=3,},
	    xticklabels={},
	    yticklabels={},
	]
        \addplot[
            matrix plot,
            fill opacity=0.5,
	     draw=white, thick,
            mesh/cols=5,
            point meta=explicit,] 
            table[x=x, y=y, meta=C] {\chamheatmap};
       \addplot[
            scatter,
            only marks,
            mark=none,
            nodes near coords,
            point meta=explicit,] 
            table[x=x, y=y, meta=C] {\chamheatmap};
	\end{axis}

	\node[fill=seablue, inode] (00) at (-0.45,5.025) {};
	\node[fill=seaorange, inode] (01) at (-0.45,3.975) {};
	\node[fill=seagreen, inode] (02) at (-0.45,2.85) {};
	\node[fill=seared, inode] (02) at (-0.45,1.675) {};
	\node[fill=seapurple, inode] (02) at (-0.45,0.5) {};

	\node[fill=seablue, inode] (10) at (0.7,-0.425) {};
	\node[fill=seaorange, inode] (01) at (2.05,-0.425) {};
	\node[fill=seagreen, inode] (02) at (3.475,-0.425) {};
	\node[fill=seared, inode] (02) at (4.825,-0.425) {};
	\node[fill=seapurple, inode] (02) at (6.225,-0.425) {};
\end{scope}

\begin{scope}[shift={(0.4315,5.1278)}]
	\node[fill=seablue, inode] (10) at (0,-0.425) {};
	\node[textnode] at (0,-0.9) {\textsf{20\%}};
	
	\node[fill=seaorange, inode] (01) at (1.25,-0.425) {};
	\node[textnode] at (1.25,-0.9) {\textsf{20.2\%}};
	
	\node[fill=seagreen, inode] (02) at (2.5,-0.425) {};
	\node[textnode] at (2.5,-0.9) {\textsf{19.9\%}};
	
	\node[fill=seared, inode] (02) at (3.75,-0.425) {};
	\node[textnode] at (3.75,-0.9) {\textsf{22.9\%}};
	
	\node[fill=seapurple, inode] (02) at (5,-0.425) {};
	\node[textnode] at (5,-0.9) {\textsf{17\%}};
\end{scope}

\begin{scope}[shift={(6.1159,5.981)}, scale=0.8]
	\begin{scope}[shift={(-9.5625,-7)}] 
		\node[inode, fill=seablue] (v1) at (9.775,3.75) {};
		\node[inode, fill=seaorange] (v2) at (10.525,3.75) {};
		\draw[iedge]  (v1) edge (v2);
		\node [textnode] at (11.4607,3.75) {\textsf{6x}};
	\end{scope}
	
	\begin{scope}[shift={(-9.3156,-7.8125)}]  
		\node[inode, fill=seaorange] (v1) at (9.25,3.75) {};
		\node[inode, fill=seaorange] (v3) at (10.75,3.75) {};
		\node[inode, fill=seablue] (v2) at (10,3.75) {};
		\draw[iedge]  (v1) edge (v2);
		\draw[iedge]  (v2) edge (v3);
		\node [textnode] at (11.237,3.75) {\textsf{4x}};
	\end{scope}

	\begin{scope}[shift={(-9.3286,-10.15)}]  
		\node[inode, fill=seagreen] (v1) at (9.25,3.8) {};
		\node[inode, fill=seagreen] (v3) at (10.75,3.8) {};
		\node[inode, fill=seagreen] (v4) at (10.75,3.25) {};
		\node[inode, fill=seagreen] (v5) at (9.25,3.25) {};
		\node[inode, fill=seared] (v2) at (10,3.55) {};
		\draw[iedge]  (v1) edge (v2);
		\draw[iedge]  (v2) edge (v3);
		\draw[iedge] (v2) edge (v4);
		\draw[iedge] (v2) edge (v5);
		\node [textnode] at (11.2896,3.5) {\textsf{3x}};
	\end{scope}

	\begin{scope}[shift={(-9.375,-11.3146)}]  
		\node[inode, fill=seaorange] (v1) at (9.25,3.75) {};
		\node[inode, fill=seablue] (v2) at (10.75,3.75) {};
		\node[inode, fill=seaorange] (v3) at (10.025,3.325) {};
		\node[inode, fill=seaorange] (v4) at (10.75,2.85) {};
		\node[inode, fill=seablue] (v5) at (9.25,2.85) {};
		
		\draw[iedge] (v1) edge (v2);
		\draw[iedge] (v2) edge (v3);
		\draw[iedge] (v3) edge (v5);
		\draw[iedge] (v2) edge (v4);
		\draw[iedge] (v4) edge (v5);
		\draw[iedge] (v5) edge (v1);
		\node [textnode] at (11.2633,3.3) {\textsf{3x}};
	\end{scope}
	
	\begin{scope}[shift={(-9.3156,-9.4271)}]  
		\node[inode, fill=seablue] (v1) at (9.25,3.75) {};
		\node[inode, fill=seablue] (v3) at (10.75,3.75) {};
		\node[inode, fill=seablue] (v2) at (10,3.75) {};
		\draw[iedge]  (v1) edge (v2);
		\draw[iedge]  (v2) edge (v3);
		\node [textnode] at (11.2633,3.75) {\textsf{3x}};
	\end{scope}
	
	\begin{scope}[shift={(-9.3286,-8.6771)}]  
		\node[inode, fill=seablue] (v1) at (9.25,3.75) {};
		\node[inode, fill=seablue] (v3) at (10.75,3.75) {};
		\node[inode, fill=seagreen] (v2) at (10,3.75) {};
		\draw[iedge]  (v1) edge (v2);
		\draw[iedge]  (v2) edge (v3);
		\node [textnode] at (11.2534,3.75) {\textsf{3x}};
	\end{scope}
\end{scope}

\end{tikzpicture}
    \caption{Chameleon. (A) Proportion of nodes in each class (color). (B) Attribute mixing matrix, each cell represents the percentage of edges with the corresponding colored endpoints. Almost uniform numbers in the matrix indicates heterophily. (C) The six most frequent rule RHS subgraphs of the extracted AVRG.}
    \label{fig:cham-rules}
\end{figure}

\para{Chameleon.} The Chameleon Wikipedia network consists of about $2500$ webpages on Wikipedia, which fall under the topic \textit{Chameleon}. 
The nodes are classified into $5$ (almost) quantiles based on their average monthly traffic, whose distribution is shown in~\autoref{fig:cham-rules}(A). 
The edge mixing patterns, in this case, is quite intriguing. 
We see that nodes lying in the same quantile are not necessarily likely to be more connected to each other than others, resulting in a more uniform distribution of edges as seen in~\autoref{fig:cham-rules}(B), resulting in the low attribute assortativity score of $0.032$.
The $6$ most frequent RHS subgraphs in the grammar rules reflect this disparity too; all but one rule consists of nodes from different classes, as seen in~\autoref{fig:cham-rules}(C). 

Thus, it is clear that the grammar rules reflect the mixing patterns observed in the input graph, which empirically validates our extraction procedure's efficacy.
Additionally, we can use the grammar rules to generate graphs which we describe next.

\section{Graph Generation with AVRGs}
The grammar $G$ encodes information about the original graph $H$, which can be used for various purposes. One compelling way to determine the faithfulness of a graph model is to use it to generate graphs. 
How closely do these generated graphs resemble the original one? Is the model able to capture the nuances of mixing patterns across both degrees and attributes?  
To answer these questions, we first describe a stochastic generative process that repeatedly applies rewriting rules to generate graphs. 

Any new graph $H^\prime$ starts with $\mathcal{S}$, a size $0$ nonterminal. 
An AVRG extracted using the extraction process outlined above is guaranteed to have a single size-$0$ starting nonterminal node because the grammar extraction only terminates when the root of the dendrogram covers all the nodes in the graph.
Then grammar rules are applied stochastically according to $f$ on available nonterminal nodes.

Precisely, each step of the generation process consists of three sequential parts: (a) choosing a nonterminal node from the current graph $H^\prime$, (b) choosing a suitable grammar rule from the grammar $G$, and (c) replacing the selected nonterminal in $H^\prime$ with the RHS subgraph of the selected rule, and rewiring it to the rest of the graph. 
This process continues until no more nonterminals remain or until some predetermined graph size is reached. 

\subsection{Selecting a Grammar Rule} 
We pick a nonterminal $X$ of size $\omega$ at \textit{random} from $H^\prime$ and find the set of grammar rules with the same LHS nonterminal, \ie, of size $\omega$.
From that set of rules, we select a rule $\mathcal{R}$ stochastically according to frequency $f$ and connect the RHS into $H^\prime$. 
This process prevents overfitting to the input graph and enables the model to flexibly generate diverse graphs that still resemble the original. 

\subsection{Choosing a Rewiring Policy}
How the RHS is connected to the existing graph matters. The degree of the chosen nonterminal in $H^\prime$ will always match its size, \ie, $\omega$. So, deleting $X$ from $H^\prime$ will break exactly $\omega$ edges (called broken edges) that then need to be rewired to nodes in the RHS.

Recall that each node in the RHS subgraph of $R$ was labeled with boundary degree $b_i$ during extraction, and the sum of all boundary degrees is equal to $\omega$. We will use this helpful property to guide the edge rewiring process. 
Doing so helps us preserve the local connectivity patterns across the different regions of the graph, thereby increasing the fidelity of the output. 
We consider three possible rewiring policies in the current work: (1) random rewiring, (2) mixing matrix, and (3) local greedy. 

\para{Random Rewiring.} Each node $i$ in the RHS subgraph of $R$ receives exactly $b_i$ edges at random from the set of broken edges.

\para{Mixing Matrix.} 
If we consider graph generation as a kind of reverse extraction process, from the top of the dendrogram moving downwards, then we can think of each RHS as introducing a new region/community of the graph--first at a macro-level and progressively refining the graph structure one rule at a time. 

As their name implies, terminal nodes and their induced subgraph stay constant once they are introduced. We, therefore, distinguish two sets of terminal nodes: (1) terminal nodes in $H^\prime$, and (2) terminal nodes in the RHS subgraph of a rule that is being applied.

From among the terminal nodes in $H^\prime$, we select those which have edges that are broken. Then from the terminal nodes in the RHS of the rewriting rule, we select terminal nodes that have non-zero boundary degrees. Once we have identified these two sets, we use a Chung-Lu style edge rewiring policy that matches edges based on the attribute mixing matrix from the input graph. 
For instance, if red nodes are twice as likely to connect to other red nodes as blue nodes in the input graph, we will use that probability to match red and blue nodes similarly during the rewiring process for the broken edges between the two terminal node sets.



\setlength{\abovedisplayskip}{3pt}
\setlength{\belowdisplayskip}{3pt}
\para{Local Greedy.} We compute an error function for each possible rewiring between two terminal nodes and greedily pick a rewiring configuration that minimizes the local error:
\begin{equation}
    \epsilon = \beta \cdot |\rho^\prime_\text{deg} - \rho_\text{deg}| + (1 - \beta) \cdot |\rho^{\prime}_\text{attr} - \rho_\text{attr}|
\end{equation}
\noindent where $\rho_\text{deg}$ and $\rho^{\prime}_\text{deg}$ is the degree assortativity before and after the application of the rewiring policy, likewise for $\rho_{\text{attr}}$. We select $\beta \in \{0, 0.5, 1\}$, which mediates the weight between the two, and call the policies Greedy-Attr, Greedy-50, and Greedy-Deg respectively.

\begin{figure}[t]
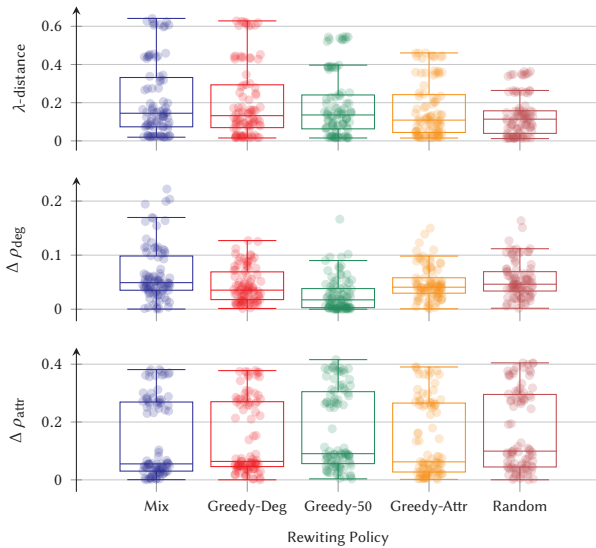

    \centering
    \iffast
        \includegraphics[height=0.42\textheight, width=0.3\textwidth]{example-image}
    \else
        \pgfplotsset{
    only if/.style args={entry of #1 is #2}{
        /pgfplots/boxplot/data filter/.code={
            \edef\tempa{\thisrow{#1}}
            \edef\tempb{#2}
            \ifx\tempa\tempb
            \else
                \def\pgfmathresult{}
            \fi
        }
    }
}

\makeatletter
\pgfplotsset{
    boxplot/hide outliers/.code={
        \def\pgfplotsplothandlerboxplot@outlier{}%
    }
}
\makeatother

\input{data/gen-type}

\begin{tikzpicture}
\begin{groupplot}[
    group style={
        group name=my plots,
        group size=1 by 3,
        xlabels at=edge bottom,
        xticklabels at=edge bottom,
        vertical sep=10pt,
    },
    width=0.48\textwidth,
    height=3.5cm,
    boxplot/draw direction=y,
    x axis line style={opacity=0},
    axis x line*=bottom,
    axis y line=left,
    enlarge y limits,
    ymajorgrids,
    xtick={1,2,3,4,5},
    xticklabels={Mix,Greedy-Deg,Greedy-50,Greedy-Attr,Random},
    x tick label style={rotate=0},
    xlabel = {Rewiting Policy},
]

    \nextgroupplot[ylabel={\textsf{$\lambda$-distance}}]
    \addplot [mark=*,only marks,blue,fill=blue,draw opacity=0.2,fill opacity=0.2,jitter=0.3,mark size=1.5] table[x=gen,y=ld]{\gengreedyattr};  
    \addplot [boxplot,  blue] table[y=ld] {\gengreedyattr};

    \addplot [mark=*,only marks,red,fill=red,draw opacity=0.2,fill opacity=0.2,jitter=0.3,mark size=1.5] table[x=gen,y=ld] {\gengreedyfifty};   
    \addplot+ [boxplot,  red] table[y=ld] {\gengreedyfifty};
    
    \addplot [mark=*,only marks,green!70!black,fill=green!70!black,draw opacity=0.2,fill opacity=0.2,jitter=0.3,mark size=1.5] table[x=gen,y=ld] {\gengreedydeg};   
    \addplot [boxplot,  green!70!black, /pgfplots/boxplot/hide outliers] table[y=ld] {\gengreedydeg};
    
    \addplot [mark=*,only marks,orange,fill=orange,draw opacity=0.2,fill opacity=0.2,jitter=0.3,mark size=1.5] table[x=gen,y=ld] {\genmix};   
    \addplot [boxplot,  orange] table[y=ld] {\genmix};
    
    \addplot [mark=*,only marks,purple,fill=purple,draw opacity=0.2,fill opacity=0.2,jitter=0.3,mark size=1.5] table[x=gen,y=ld] {\genrandom};   
    \addplot [boxplot,  purple, /pgfplots/boxplot/hide outliers] table[y=ld] {\genrandom};
    
    
    
    \nextgroupplot[ylabel={$\Delta$ $\rho_{\textsf{deg}}$}]
    
    \addplot [mark=*,only marks,blue,fill=blue,draw opacity=0.2,fill opacity=0.2,jitter=0.3,mark size=1.5] table[x=gen,y=deg]{\gengreedyattr};  
    \addplot [boxplot,  blue, /pgfplots/boxplot/hide outliers] table[y=deg] {\gengreedyattr};

    \addplot [mark=*,only marks,red,fill=red,draw opacity=0.2,fill opacity=0.2,jitter=0.3,mark size=1.5] table[x=gen,y=deg] {\gengreedyfifty};   
    \addplot [boxplot,  red] table[y=deg] {\gengreedyfifty};

    \addplot [mark=*,only marks,green!70!black,fill=green!70!black,draw opacity=0.2,fill opacity=0.2,jitter=0.3,mark size=1.5] table[x=gen,y=deg] {\gengreedydeg};   
    \addplot [boxplot,  green!70!black, /pgfplots/boxplot/hide outliers] table[y=deg] {\gengreedydeg};

    \addplot [mark=*,only marks,orange,fill=orange,draw opacity=0.2,fill opacity=0.2,jitter=0.3,mark size=1.5] table[x=gen,y=deg] {\genmix};   
    \addplot [boxplot,  orange, /pgfplots/boxplot/hide outliers] table[y=deg] {\genmix};

    \addplot [mark=*,only marks,purple,fill=purple,draw opacity=0.2,fill opacity=0.2,jitter=0.3,mark size=1.5] table[x=gen,y=deg] {\genrandom};   
    \addplot [boxplot,  purple, /pgfplots/boxplot/hide outliers] table[y=deg] {\genrandom};


    \nextgroupplot[ylabel={$\Delta$ $\rho_{\textsf{attr}}$}]

    \addplot [mark=*,only marks,blue,fill=blue,draw opacity=0.2,fill opacity=0.2,jitter=0.3,mark size=1.5] table[x=gen,y=attr]{\gengreedyattr};  
    \addplot [boxplot,  blue] table[y=attr] {\gengreedyattr};

    \addplot [mark=*,only marks,red,fill=red,draw opacity=0.2,fill opacity=0.2,jitter=0.3,mark size=1.5] table[x=gen,y=attr] {\gengreedyfifty};   
    \addplot [boxplot,  red] table[y=attr] {\gengreedyfifty};

    \addplot [mark=*,only marks,green!70!black,fill=green!70!black,draw opacity=0.2,fill opacity=0.2,jitter=0.3,mark size=1.5] table[x=gen,y=attr] {\gengreedydeg};   
    \addplot [boxplot,  green!70!black] table[y=attr] {\gengreedydeg};

    \addplot [mark=*,only marks,orange,fill=orange,draw opacity=0.2,fill opacity=0.2,jitter=0.3,mark size=1.5] table[x=gen,y=attr] {\genmix};   
    \addplot [boxplot,  orange] table[y=attr] {\genmix};

    \addplot [mark=*,only marks,purple,fill=purple,draw opacity=0.2,fill opacity=0.2,jitter=0.3,mark size=1.5] table[x=gen,y=attr] {\genrandom};   
    \addplot [boxplot,  purple] table[y=attr] {\genrandom};
\end{groupplot}
\end{tikzpicture}
    \fi
    \caption{Rewiring policies over three metrics. Lower is better. We select the Mixing Matrix rewiring policy.}
    \label{fig:rewiring}
\end{figure}

\para{Selecting a Rewiring Policy.} Among the many rewiring policies available to our model, it is crucial to choose one that generates graphs that contain mixing patterns that are faithful to the original graph. Based on the results from previous experiments, we choose the Leiden clustering and a size-based node selection strategy with $\mu=5$. Then we test the five rewiring policies with ten repetitions on each dataset in Table~\ref{tab:datasets}. 

Figure~\ref{fig:rewiring} illustrates the individual results for each trial and a corresponding box and whisker plot for $\lambda$-distance~\citep{wilson2008study}, $\Delta\rho_\text{deg}$, and $\Delta\rho_\text{attr}$ \ie, absolute difference between the degree and attribute assortativity of the generated and original graphs, respectively (lower is better). Given the lack of a clear winner among the rewiring policies, we choose the mixing matrix one in the remainder of this work, given its flexibility.

\begin{figure}[t]
    \centering
    \iffast
        \includegraphics[height=0.31\textheight, width=0.4\textwidth]{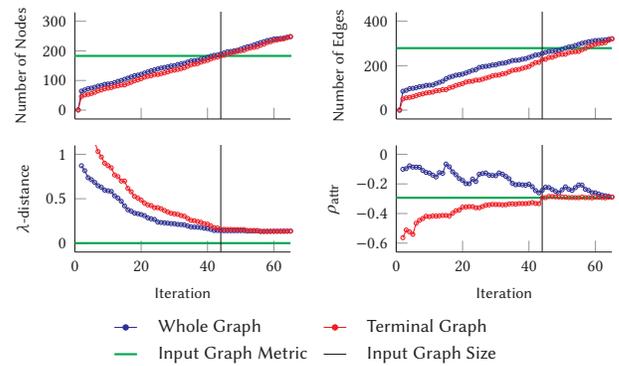}
    \else
        \pgfplotstableread{
t   n   m   ld  attr
1	1	0   nan nan
2	64	85	0.87156	-0.10097144539299385
3	68	89	0.81546	-0.09683098591549287
4	72	96	0.73694	-0.07775164985208216
5	74	98	0.7145	-0.08674399941635662
6	77	101	0.68363	-0.08653264304444298
7	81	105	0.64616	-0.08747919363787679
8	83	107	0.62878	-0.10887357946181471
9	87	111	0.59675	-0.1220652173913043
10	88	112	0.58911	-0.12713452170654688
11	89	113	0.58169	-0.14051411449298215
12	93	120	0.53248	-0.15293129984608936
13	97	125	0.50693	-0.12395632626846495
14	100	130	0.471	-0.12816062482742302
15	104	141	0.4192	-0.06578815344994818
16	108	148	0.37539	-0.0846431059023045
17	112	152	0.35972	-0.11800028731504104
18	116	157	0.33778	-0.1381119826886665
19	118	159	0.33127	-0.163792422527053
20	121	162	0.32189	-0.176369733491339
21	125	167	0.30981	-0.1978149863436646
22	129	173	0.28455	-0.19863195057369812
23	133	179	0.27337	-0.16755891945111065
24	136	182	0.26657	-0.18386974479717458
25	139	192	0.23764	-0.1350629462775334
26	142	195	0.2321	-0.12998637531690319
27	144	197	0.2285	-0.12662315817802594
28	146	199	0.22502	-0.14120923487632459
29	149	203	0.21953	-0.14235299728170156
30	150	204	0.21788	-0.14424364650459282
31	154	209	0.21118	-0.12629672320105395
32	156	211	0.20817	-0.13422432252824337
33	158	213	0.20517	-0.14863082208660117
34	162	218	0.19346	-0.16490392602026688
35	166	223	0.18266	-0.19572632556904906
36	167	224	0.1814	-0.20587420338043794
37	168	225	0.18016	-0.20377655389457114
38	172	229	0.17546	-0.20643845851854664
39	174	232	0.16784	-0.20870628032444838
40	178	237	0.16386	-0.20210908747758813
41	182	242	0.15418	-0.2187546288383752
42	185	248	0.14238	-0.24496633308960727
43	186	249	0.14234	-0.2602939100022809
44	189	255	0.13971	-0.24343981156246128
45	193	260	0.13957	-0.22235720564559927
46	197	264	0.13954	-0.2243743324769795
47	198	265	0.13954	-0.2380450911875928
48	200	267	0.13964	-0.24964944859210964
49	202	269	0.13982	-0.2618326125094989
50	206	274	0.14021	-0.24020871314515635
51	210	279	0.14047	-0.22021816052964652
52	214	284	0.13567	-0.23622414943295542
53	216	286	0.13587	-0.24436973127981995
54	220	291	0.13646	-0.22513710500700368
55	224	296	0.13708	-0.20725313691323183
56	228	301	0.13267	-0.20829121069681297
57	231	304	0.133	-0.23657411119777824
58	235	308	0.1334	-0.24420463629096714
59	237	310	0.13378	-0.25949362608738735
60	240	313	0.13427	-0.2602393580096078
61	241	314	0.13436	-0.2681955240808873
62	242	315	0.13451	-0.2763230429988976
63	243	316	0.13468	-0.28017797797247934
64	246	319	0.13517	-0.28411246727637446
65	248	321	0.13556	-0.28871226099474856
}{\growthwhole}

\pgfplotstableread{
t	n	m	ld	attr
1	0	0   nan nan
2	46	50	1.58285	-0.5630427231677668
3	51	55	1.40475	-0.5114042319318494
4	53	57	1.33323	-0.5235158057054743
5	56	60	1.24274	-0.5406162464985994
6	60	65	1.13694	-0.4658898703251268
7	65	70	1.03115	-0.44605116796440497
8	68	73	0.97002	-0.43406037844864964
9	72	77	0.89983	-0.4186284544524051
10	74	79	0.87258	-0.42117418256958783
11	76	81	0.84742	-0.4166379012771835
12	80	87	0.76744	-0.4170225444433409
13	82	88	0.74848	-0.41619001454192917
14	85	92	0.69896	-0.4130108570924176
15	85	92	0.69896	-0.4130108570924176
16	90	100	0.61599	-0.4165361962275405
17	94	104	0.57789	-0.39779194953027497
18	99	111	0.5262	-0.3807137250166054
19	102	115	0.50715	-0.37670619825464313
20	106	119	0.48499	-0.3567870165887731
21	111	126	0.45097	-0.35601258607323627
22	114	129	0.43028	-0.35394890922496425
23	116	130	0.42258	-0.3539932556451957
24	120	135	0.3989	-0.3610923669273211
25	120	135	0.3989	-0.3610923669273211
26	123	139	0.38725	-0.35269678481865446
27	126	144	0.36747	-0.34360852887306914
28	129	148	0.35073	-0.34190013372036143
29	133	154	0.33587	-0.33820112844341166
30	135	157	0.32953	-0.3392717249383431
31	137	158	0.32459	-0.33944323519303454
32	140	163	0.30978	-0.3414911473279549
33	143	165	0.30297	-0.3408263416178567
34	148	172	0.27872	-0.3354874490607895
35	153	179	0.25638	-0.3349610458792978
36	155	181	0.25252	-0.33648959735916256
37	157	183	0.24335	-0.33312032717745554
38	162	188	0.23429	-0.3330915976712797
39	165	192	0.21868	-0.3332047449127208
40	168	196	0.21399	-0.3312415791513629
41	173	206	0.19411	-0.32668179391350766
42	177	214	0.17291	-0.331682172102599
43	179	215	0.17085	-0.33300593913537235
44	183	228	0.15574	-0.2924093650516476
45	185	229	0.15563	-0.2929755642852387
46	190	237	0.15412	-0.2845368770433723
47	192	238	0.15405	-0.2852017605503569
48	195	244	0.1526	-0.2866936192652486
49	198	250	0.15224	-0.2882573391630232
50	200	251	0.15225	-0.2888460985137722
51	202	252	0.15227	-0.28944132115249466
52	207	259	0.14226	-0.29238829864648536
53	210	262	0.14218	-0.29191059343959574
54	212	263	0.14225	-0.29250042800028203
55	214	264	0.14233	-0.29309569267985486
56	219	272	0.13109	-0.2887965478663898
57	223	278	0.13106	-0.29266204684106645
58	228	286	0.13145	-0.29227516475272175
59	231	291	0.13197	-0.2941748132530617
60	235	297	0.13267	-0.28952688087466105
61	237	299	0.13291	-0.2907643930276796
62	239	301	0.13287	-0.29209305890920256
63	241	304	0.13335	-0.2938018362144809
64	245	314	0.13425	-0.2892820860142228
65	248	321	0.13556	-0.28871226099474856
}{\growthterminal}

\begin{tikzpicture}
\begin{groupplot}[
    log ticks with fixed point,
    group style={
        group name=my plots,
        group size=2 by 2,
        xlabels at=edge bottom,
        xticklabels at=edge bottom,
        vertical sep=10pt,
        horizontal sep=40pt,
    },
    width=0.25\textwidth,
    height=3cm,
    enlarge y limits,
    xlabel={Iteration}, 
    xmin=0, xmax=65,
    xtick pos=bottom,
    ytick pos=left,
    axis x line*=bottom,
    axis y line*=left,
]

\nextgroupplot[ylabel={Number of Nodes}, ymin=0, ymax = 300]
\addplot [mark=o, color=blue, mark size=0.75] table[x=t,y=n]{\growthwhole}; 
\addplot [mark=o, color=red, mark size=0.75] table[x=t,y=n]{\growthterminal}; 
\addplot[color=green, thick, no markers] coordinates {(-10, 183) (165, 183)};
\addplot[color=black, no markers] coordinates {(44, -100) (44, 500)};

\nextgroupplot[ylabel={Number of Edges}, ymin=0, ymax = 400]
\addplot [mark=o, color=blue, mark size=0.75] table[x=t,y=m]{\growthwhole}; 
\addplot [mark=o, color=red, mark size=0.75] table[x=t,y=m]{\growthterminal}; 
\addplot[color=green, thick, no markers] coordinates {(0, 279) (65,279)}; 
\addplot[color=black, no markers] coordinates {(44, -100) (44, 500)};

\nextgroupplot[ylabel={$\lambda$-distance}, ymax=1, ymin=0]
\addplot [mark=o, color=blue, mark size=0.75] table[x=t,y=ld]{\growthwhole}; 
\addplot [mark=o, color=red, mark size=0.75] table[x=t,y=ld]{\growthterminal}; 
\addplot[color=green, thick, no markers] coordinates {(0, 0) (65, 0)};
\addplot[color=black, no markers] coordinates {(44, -1) (44, 2)};

\nextgroupplot[ylabel={$\rho_{\textsf{attr}}$}, ymax=0, ymin=-0.6]
\addplot [mark=o, color=blue, mark size=0.75] table[x=t,y=attr]{\growthwhole}; 
\addplot [mark=o, color=red, mark size=0.75] table[x=t,y=attr]{\growthterminal}; 
\addplot[color=green, thick, no markers] coordinates {(0, -0.2936439815593172) (65, -0.2936439815593172)};
\addplot[color=black, no markers] coordinates {(44, -1) (44, 0.1)};

\end{groupplot}
\end{tikzpicture}

\begin{tikzpicture}
    \begin{customlegend}[ 
    legend columns=2,
    legend style={
    draw=none,
    font=\footnotesize,
    column sep=1ex,
  },
  legend cell align={left},
  legend entries={~\textsf{Whole Graph}~,~\textsf{Terminal Graph}~,~\textsf{Input Graph Metric}~, ~\textsf{Input Graph Size}~}
  ]
    \addlegendimage{draw=blue, mark=*, fill=blue, mark size=1}
    \addlegendimage{draw=red, mark=*, fill=red, mark size=1}
    \addlegendimage{draw=green, thick, no markers}
    \addlegendimage{draw=black, no markers}
    \end{customlegend}
\end{tikzpicture}
    \fi
    
    \caption{The growth of an example graph (Texas) during generation. The Horizontal green line indicates the corresponding value in the original graph. The generated graph can be made to grow to an arbitrary size; the vertical black line represents the size of the input graph. This example shows how the metrics of the generated graphs slowly approach and eventually converge near ideal value as the generated graph approaches the original graph's size.}
    \label{fig:temporal-growth}
\end{figure}

\subsection{Observing the Generation Dynamics}

Next, we inspect the graph generation process. Observing the growth of the generator provides a unique look at its underlying dynamics. Despite the temptation, it is important to \textit{not} think of this generation process as a temporal one, but rather as a top-down filling-out. As new nodes and edges are added, the size of the graph grows, and we expect that the underlying topological and attribute mixtures will converge near the correct values.

Four views of graph generation on the Texas graph are shown in~\autoref{fig:temporal-growth}. In this example, the AVRG was extracted from the Texas graph using Leiden clustering. Rules were extracted using the size-based scoring policy ($\mu=5$), breaking ties randomly. In this example, the graph generator used the mixing matrix rewiring policy and was requested to grow to 250 nodes instead of 183, \ie, we intentionally specified the new graph to be larger than the original. The difference between blue and red points shows the difference between the entire generated graph (treating terminal and nonterminal nodes as the same) versus just the terminal graph (ignoring nonterminal nodes and edges adjacent to nonterminal nodes). These values will eventually converge. When there are no more nonterminals, graph generation ceases.

Green horizontal lines in \autoref{fig:temporal-growth} show the graph metrics from the original graph (except $\lambda$-distance, where 0 means perfect match). In this example, we asked the generative model to produce a graph with an extra 67 nodes. As rewriting rules are applied, the number of nodes gradually approaches 250. Likewise, the graph metrics converge to near the values of the input graph. The graph metrics remain relatively stable once the input size is reached--indicated by the vertical black line. 

This example illustrates an important property of AVRGs: they can generate graphs of any large size, yet they continue to resemble the input graph.

\begin{table}[t]
    \centering
    \caption{Dataset summary statistics. $|V|$: \# of nodes, $|E|$: \# of edges, $|\mathcal{A}|$: \# of unique attribute values, $\rho_{\text{deg}}$: degree assortativity coefficient, $\rho_{\text{attr}}$: attribute assortativity coefficient.}
    \small{
    \begin{tabular}{@{}lrrrrr@{}}
        \toprule
        \textbf{Name} &  $\mathbf{|V|}$ &    $\mathbf{|E|}$ &  $|\mathbf{\mathcal{A}}|$ &  $\mathbf{\rho_{\text{deg}}}$ &  $\mathbf{\rho_{\text{attr}}}$  \\
        \midrule
        Polbooks &  105 &    441 &    3 & $-$0.128 &  0.723 \\
        Football &  115 &    613 &   12 &  0.162 &  0.608 \\
        Texas &  183 &    279 &    5 & $-$0.270 & $-$0.294 \\
        Cornell &  183 &    277 &    5 & $-$0.249 & $-$0.079 \\
        Wisconsin &  251 &    450 &    5 & $-$0.193 & $-$0.173 \\
        Airports & 1,183 &  22,459 &    4 &  0.168 &  0.356\\
        Polblogs & 1,222 &  16,714 &    2 & $-$0.221 &  0.811 \\
        Citeseer & 2,110 &   3,668 &    6 &  0.007 &  0.664 \\
        Chameleon & 2,277 &  31,371 &    5 & $-$0.200 &  0.032 \\
        Cora & 2,485 &   5,069 &    7 & $-$0.071 &  0.764 \\
        Squirrel & 5,201 & 198,353 &    5 & $-$0.227 &  0.007\\
        Film & 7,600 &  26,659 &    5 & $-$0.047 &  0.003 \\
        \bottomrule
    \end{tabular}
    }
    \label{tab:datasets}
\end{table}

\section{Graph Generation Results and Findings} 
We performed a thorough set of experiments that compared the graph generation performance of the AVRG model against several state-of-the-art graph models. Note that, just like in other graph grammar models, applying grammar rules in a tree-ordering over the dendrogram will permit the generation of an isomorphic graph to the input graph. However, this property of AVRGs is not particularly helpful in generating null graphs. Instead, our goal was to generate new graphs and determine if these new graphs shared similar topological and attribute properties with the input graphs.

\begin{figure}[tb]
    \centering
    \vspace{0.5cm}
    \iffast
        \includegraphics[height=0.26\textheight, width=0.3\textwidth]{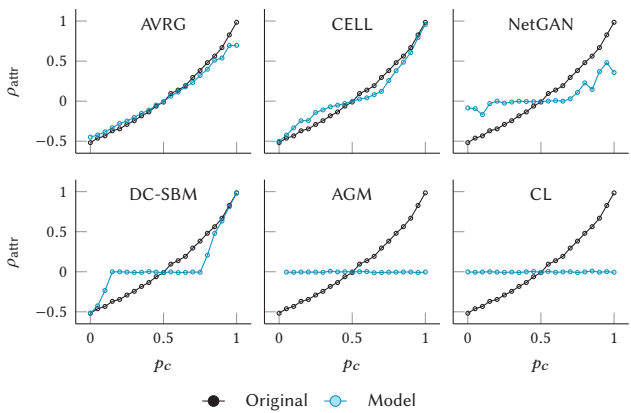}
    \else
        \pgfplotstableread{
pc  model  attr
0.0 orig -0.517691817601719
0.05 orig -0.46171737376968514
0.1 orig -0.4332678256944926
0.15000000000000002 orig -0.36977355839383147
0.2 orig -0.3459899425396446
0.25 orig -0.29067736058219484
0.30000000000000004 orig -0.24429373815099617
0.35000000000000003 orig -0.1831384039270923
0.4 orig -0.1345591134784107
0.45 orig -0.062012095773266586
0.5 orig -0.011017332102994812
0.55 orig 0.09492528965886272
0.6000000000000001 orig 0.1382927831422776
0.65 orig 0.1925910029966904
0.7000000000000001 orig 0.2954534572936214
0.75 orig 0.3820733354627461
0.8 orig 0.4804035029048454
0.8500000000000001 orig 0.562272235817129
0.9 orig 0.6684428852332902
0.9500000000000001 orig 0.8273071828599949
1.0 orig 0.9845685446929624
}{\cabamorig}

\pgfplotstableread{
pc  model  attr
0 AVRG -0.449
0.05 AVRG -0.422
0.1 AVRG -0.384
0.15 AVRG -0.332
0.2 AVRG -0.28
0.25 AVRG -0.248
0.3 AVRG -0.203
0.35 AVRG -0.152
0.4 AVRG -0.113
0.45 AVRG -0.0492
0.5 AVRG -0.00939
0.55 AVRG 0.0633
0.6 AVRG 0.114
0.65 AVRG 0.179
0.7 AVRG 0.232
0.75 AVRG 0.321
0.8 AVRG 0.399
0.85 AVRG 0.512
0.9 AVRG 0.538
0.95 AVRG 0.694
1 AVRG 0.697
}{\cabamavrg}

\pgfplotstableread{
pc  model  attr
0.05 AGM -0.0045
0.1 AGM -0.007
0.15 AGM -0.0038
0.2 AGM -0.004
0.25 AGM -0.0046
0.3 AGM -0.0074
0.35 AGM 0.0072
0.4 AGM -0.0027
0.45 AGM 0.0002
0.5 AGM -0.0003
0.55 AGM 0.0006
0.6 AGM 0.0024
0.65 AGM -0.0125
0.7 AGM -0.0092
0.75 AGM -0.0058
0.8 AGM -0.0054
0.85 AGM -0.0035
0.9 AGM -0.0032
0.95 AGM -0.0105
1.0 AGM -0.0036
}{\cabamagm}

\pgfplotstableread{
pc  model  attr
0.0 CL -0.0016
0.05 CL -0.0063
0.1 CL -0.0035
0.15 CL 0.0
0.2 CL -0.0043
0.25 CL -0.0084
0.3 CL -0.0038
0.35 CL -0.0097
0.4 CL -0.0011
0.45 CL 0.0033
0.5 CL -0.0076
0.55 CL 0.0044
0.6 CL -0.0071
0.65 CL -0.0025
0.7 CL -0.0002
0.75 CL -0.0098
0.8 CL -0.002
0.85 CL 0.0087
0.9 CL -0.0057
0.95 CL 0.0008
1.0 CL -0.0064
}{\cabamcl}

\pgfplotstableread{
pc  model  attr
0.0 DC-SBM -0.5185
0.05 DC-SBM -0.4253
0.1 DC-SBM -0.2345
0.15 DC-SBM -0.0014
0.2 DC-SBM -0.0006
0.25 DC-SBM -0.004
0.3 DC-SBM -0.0091
0.35 DC-SBM -0.0081
0.4 DC-SBM 0.0019
0.45 DC-SBM -0.0044
0.5 DC-SBM -0.0118
0.55 DC-SBM -0.0018
0.6 DC-SBM -0.0111
0.65 DC-SBM -0.0074
0.7 DC-SBM -0.0034
0.75 DC-SBM -0.0063
0.8 DC-SBM 0.2063
0.85 DC-SBM 0.4802
0.9 DC-SBM 0.6331
0.95 DC-SBM 0.8126
1.0 DC-SBM 0.9827
}{\cabamdcsbm}

\pgfplotstableread{
pc  model  attr
0.0 netgan -0.0849
0.05 netgan -0.0938
0.1 netgan -0.1687
0.15 netgan -0.0299
0.2 netgan -0.0004
0.25 netgan -0.0253
0.3 netgan -0.0114
0.35 netgan -0.0002
0.4 netgan -0.0048
0.45 netgan -0.0075
0.5 netgan -0.0042
0.55 netgan 0.0003
0.6 netgan 0.0063
0.65 netgan -0.0005
0.7 netgan 0.0291
0.75 netgan 0.1091
0.8 netgan 0.2273
0.85 netgan 0.1465
0.9 netgan 0.3678
0.95 netgan 0.4805
1.0 netgan 0.3569
}{\cabamnetgan}

\pgfplotstableread{
pc  model  attr
0.0 cell -0.5015
0.05 cell -0.424
0.1 cell -0.3332
0.15 cell -0.245
0.2 cell -0.2433
0.25 cell -0.1407
0.3 cell -0.1098
0.35 cell -0.0701
0.4 cell -0.0502
0.45 cell -0.0293
0.5 cell -0.0045
0.55 cell 0.026
0.6 cell 0.041
0.65 cell 0.0813
0.7 cell 0.1216
0.75 cell 0.2571
0.8 cell 0.3794
0.85 cell 0.4867
0.9 cell 0.6061
0.95 cell 0.7928
1.0 cell 0.9582
}{\cabamcell}
\begin{tikzpicture}
\begin{groupplot}[
    log ticks with fixed point,
    group style={
        group name=my plots,
        group size=3 by 2,
        x descriptions at=edge bottom,
        y descriptions at =edge left,
        vertical sep=10pt,
        horizontal sep=5pt,
    },
    ymin=-0.5,
    ymax=1,
    xmin=0,
    xmax=1,
    width=0.22\textwidth,
    height=3.5cm,
    enlargelimits=0.1,
    xlabel={$p_c$},
    ylabel={$\rho_{\textsf{attr}}$},
    title style = {yshift=-17pt},
    xtick pos=bottom,
    ytick pos=left,
    axis x line*=bottom,
    axis y line*=left,
]

\nextgroupplot[title={AVRG}]
\addplot [mark=o, color=black, mark size=0.75] table[x=pc,y=attr]{\cabamorig}; 
\addplot [mark=o,color=cyan!70!black,mark size=0.75] table[x=pc,y=attr]{\cabamavrg}; 


\nextgroupplot[title={CELL}]
\addplot [mark=o, color=black, mark size=0.75] table[x=pc,y=attr]{\cabamorig}; 
\addplot [mark=o,color=cyan!70!black,mark size=0.75] table[x=pc,y=attr]{\cabamcell};

\nextgroupplot[title={NetGAN}]
\addplot [mark=o, color=black, mark size=0.75] table[x=pc,y=attr]{\cabamorig}; 
\addplot [mark=o,color=cyan!70!black, mark size=0.75] table[x=pc,y=attr]{\cabamnetgan};

\nextgroupplot[title={DC-SBM}]
\addplot [mark=o, color=black, mark size=0.75] table[x=pc,y=attr]{\cabamorig}; 
\addplot [mark=o,color=cyan!70!black, mark size=0.75] table[x=pc,y=attr]{\cabamdcsbm}; 

\nextgroupplot[title={AGM}]
\addplot [mark=o, color=black, mark size=0.75] table[x=pc,y=attr]{\cabamorig}; 
\addplot [mark=o,color=cyan!70!black, mark size=0.75] table[x=pc,y=attr]{\cabamagm}; 

\nextgroupplot[title={CL}]
\addplot [mark=o, color=black, mark size=0.75] table[x=pc,y=attr]{\cabamorig}; 
\addplot [mark=o,color=cyan!70!black, mark size=0.75] table[x=pc,y=attr]{\cabamcl};

\end{groupplot}
\end{tikzpicture} 

\begin{tikzpicture}
    \begin{customlegend}[ 
    legend columns=2,
    legend style={
    draw=none,
    font=\footnotesize,
    column sep=1ex,
  },
  legend cell align={left},
  legend entries={~\textsf{Original}~,~\textsf{Model}~}
  ]
    \addlegendimage{mark=*, color=black,scale=0.8}
    \addlegendimage{mark=*,draw=cyan!70!black,fill=cyan!30!white,scale=0.8}
    \end{customlegend}
\end{tikzpicture}
    \fi
    \caption{Results on matching simulated scale-free graphs: AVRG best hugs the assortativity curve (CELL is a close contender) across the spectrum, while other models struggle.}
    \label{fig:cabam}
\end{figure}

\begin{table*}[tb]
\caption{Graph generation performance. Lower is better. 95\% confidence intervals are not shown for clarity. The best and second best performing models are marked in boldface and underline respectively. Failures are marked with --. $\lambda$-dist: topological similarity, $\Delta \rho_{\text{deg}}$ and $\Delta \rho_{\text{attr}}$: difference in degree and attribute assortativities. 1 - $r$: Inverse Pearson Correlation Coefficient of colored graphlet counts. AVRG ranks in the top-2 for at least one metric across all datasets except Cora.}
    \label{tab:topology}
    {\footnotesize
    \begin{tabular}{@{}llrrrrrrrrrrrrrrr@{}}
\toprule
{}       & \multicolumn{4}{c}{Polbooks} & \multicolumn{4}{c}{Football}  & \multicolumn{4}{c}{Texas} & \multicolumn{4}{c}{Cornell}   \\ 
\cmidrule(lr){2-5} \cmidrule(lr){6-9} \cmidrule(lr){10-13} \cmidrule(ll){14-17}  
    & $\lambda$-dist          & $\Delta \rho_{\text{deg}}$          & $\Delta \rho_{\text{attr}}$         & 1 - $r$     & $\lambda$-dist          & $\Delta \rho_{\text{deg}}$          & $\Delta \rho_{\text{attr}}$         & 1 - $r$     & $\lambda$-dist          & $\Delta \rho_{\text{deg}}$          & $\Delta \rho_{\text{attr}}$         & 1 - $r$     & $\lambda$-dist          & $\Delta \rho_{\text{deg}}$          & $\Delta \rho_{\text{attr}}$         & 1 - $r$     \\ \midrule
AGM      & 0.087          & 0.062    & \textbf{0.058} & 0.165          & --             & --             & --             & --             & --             & --             & --             & --             & --             & --             & --             & --             \\
CL       & \ul{0.069}    & 0.075          & 0.729          & 1.172          & 0.178          & 0.176          & 0.618          & 1.192          & 0.265          & 0.097          & 0.276          & 0.894          & 0.226          & 0.091          & 0.058          & 0.723          \\
DC-SBM   & 0.194          & \ul{0.037}          & 0.115          & 1.154          & 0.254          & 0.142          & 0.197          & 1.129          & 0.221          & 0.051          & 0.253          & 1.366          & 0.208          & 0.056          & 0.052          & 1.305          \\
NetGAN   & 0.192          & \textbf{0.030} & 0.112          & \textbf{0.101} & 0.189          & 0.159          & 0.170          & 0.261          & 0.203          & 0.049          & 0.173          & 0.822          & 0.153          & \ul{0.023}    & \ul{0.037}    & 0.453          \\
GAE  & 3.014          & 0.093          & 0.280          & --             & 2.430          & 0.208          & 0.444          & --             & 5.584          & 0.592          & 0.326          & --             & 5.916          & 0.671          & 0.062          & --             \\
GVAE & 3.826          & 0.081          & 0.259          & --             & 3.454          & 0.199          & 0.503          & --             & 6.433          & 0.754          & 0.302          & --             & 6.235          & 0.754          & 0.093          & --             \\
CELL     & \textbf{0.058} & 0.164          & \ul{0.069}    & 0.161          & \ul{0.125}    & \ul{0.086}    & \ul{0.103}    & \ul{0.198}    & \ul{0.132}    & \ul{0.038}    & \ul{0.048}    & \ul{0.591}    & \ul{0.085}    & \textbf{0.021} & \textbf{0.032} & \textbf{0.234} \\
\rowcolor{gray!10}
AVRG     & 0.229          & 0.080          & 0.171          & \ul{0.148}    & \textbf{0.051} & \textbf{0.056} & \textbf{0.022} & \textbf{0.075} & \textbf{0.118} & \textbf{0.036} & \textbf{0.030} & \textbf{0.504} & \textbf{0.081} & 0.031          & 0.042          & \ul{0.346}    \\
\\ 

{}       & \multicolumn{4}{c}{Wisconsin}                                     & \multicolumn{4}{c}{Airports}                                      & \multicolumn{4}{c}{Polblogs}                                      & \multicolumn{4}{c}{Citeseer}                                      \\ 
\cmidrule(lr){2-5} \cmidrule(lr){6-9} \cmidrule(lr){10-13} \cmidrule(ll){14-17}  
    & $\lambda$-dist          & $\Delta \rho_{\text{deg}}$          & $\Delta \rho_{\text{attr}}$         & 1 - $r$     & $\lambda$-dist          & $\Delta \rho_{\text{deg}}$          & $\Delta \rho_{\text{attr}}$         & 1 - $r$     & $\lambda$-dist          & $\Delta \rho_{\text{deg}}$          & $\Delta \rho_{\text{attr}}$         & 1 - $r$     & $\lambda$-dist          & $\Delta \rho_{\text{deg}}$          & $\Delta \rho_{\text{attr}}$         & 1 - $r$     \\ \midrule
AGM      & --             & --             & --             & --             & \ul{0.185}    & 0.211          & \ul{0.048}    & \textbf{0.021} & 0.224          & 0.142          & \textbf{0.024} & \ul{0.006}    & 0.022          & \textbf{0.021} & \textbf{0.012} & \textbf{0.054} \\
CL       & 0.195          & 0.070          & 0.147          & 0.748          & \textbf{0.154} & 0.226          & 0.359          & 0.654          & \ul{0.150}    & 0.159          & 0.814          & 1.125          & \textbf{0.015} & 0.026          & 0.669          & 1.050          \\
DC-SBM   & 0.150          & \ul{0.035}    & 0.139          & 1.137          & 0.523          & \ul{0.025}    & \ul{0.048}    & \ul{0.650}    & 0.291          & \textbf{0.023} & 0.032          & 1.220          & \ul{0.018}    & \ul{0.022}    & 0.314          & 1.023          \\
NetGAN   & 0.198          & \ul{0.035}    & 0.115          & 0.461          & 0.589          & 0.148          & 0.152          & \textbf{0.021} & 0.323          & 0.154          & 0.367          & 0.186          & 0.032          & 0.051          & 0.602          & 0.545          \\
GAE  & 6.595          & 0.451          & 0.182          & --             & 13.518         & 0.029          & 0.154          & --             & 16.025         & 0.167          & 0.218          & --             & 20.260         & 0.027          & 0.612          & --             \\
GVAE & 7.216          & 0.547          & 0.190          & --             & 13.685         & 0.226          & 0.183          & --             & 16.079         & 0.142          & 0.222          & --             & 20.764         & 0.059          & 0.609          & --             \\
CELL     & \textbf{0.111} & \textbf{0.025} & \ul{0.029}    & \textbf{0.110} & 0.194          & \textbf{0.011} & \textbf{0.038} & \ul{0.650}    & \textbf{0.107} & \ul{0.026}    & \ul{0.028}    & \textbf{0.003} & 0.024          & 0.058          & \ul{0.200}    & \ul{0.073}    \\
\rowcolor{gray!10}
AVRG     & \ul{0.136}    & \textbf{0.025} & \textbf{0.020} & \ul{0.244}    & 0.447          & 0.041          & 0.063          & \textbf{0.021} & 0.357          & 0.083          & 0.283          & 0.123          & \ul{0.018}    & 0.034          & 0.263          & 0.101          \\
\\ 

{}       & \multicolumn{4}{c}{Chameleon}                                     & \multicolumn{4}{c}{Cora}                                          & \multicolumn{4}{c}{Squirrel}                                      & \multicolumn{4}{c}{Film}                                          \\ 
\cmidrule(lr){2-5} \cmidrule(lr){6-9} \cmidrule(lr){10-13} \cmidrule(ll){14-17}  
    & $\lambda$-dist          & $\Delta \rho_{\text{deg}}$          & $\Delta \rho_{\text{attr}}$         & 1 - $r$     & $\lambda$-dist          & $\Delta \rho_{\text{deg}}$          & $\Delta \rho_{\text{attr}}$         & 1 - $r$     & $\lambda$-dist          & $\Delta \rho_{\text{deg}}$          & $\Delta \rho_{\text{attr}}$         & 1 - $r$     & $\lambda$-dist          & $\Delta \rho_{\text{deg}}$          & $\Delta \rho_{\text{attr}}$         & 1 - $r$     \\ \midrule
AGM      & 0.196          & 0.125          & 0.036          & --             & 0.026          & 0.032    & \textbf{0.013} & \ul{0.255}    & 0.521          & 0.083          & 0.008          & --             & 0.047          & 0.020          & \ul{0.004}    & --             \\
CL       & \ul{0.174}    & 0.129          & 0.035          & --             & \textbf{0.015} & 0.046          & 0.766          & 1.114          & 0.472          & 0.095          & 0.007          & --             & 0.039          & 0.019          & \textbf{0.003} & --             \\
DC-SBM   & 0.361          & 0.017          & \ul{0.012}    & --             & 0.024          & \ul{0.020} & 0.292          & 1.078          & 0.835          & \ul{0.034}    & \ul{0.002}    & --             & 0.030          & 0.008          & \ul{0.004}    & --             \\
NetGAN   & 0.332          & 0.129          & 0.029          & --             & 0.058          & 0.036          & 0.710          & 1.089          & 1.041          & 0.219          & 0.008          & --             & --             & --             & --             & --             \\
GAE  & --             & --             & --             & --             & 22.364         & 0.066          & 0.688          & --             & --             & --             & --             & --             & --             & --             & --             & --             \\
GVAE & --             & --             & --             & --             & 21.860         & 0.058          & 0.686          & --             & --             & --             & --             & --             & --             & --             & --             & --             \\ 
CELL     & 0.260          & \textbf{0.003} & 0.017          & --             & \ul{0.022}    & \textbf{0.004}          & \ul{0.132}    & \textbf{0.067} & \textbf{0.303} & \textbf{0.030} & \textbf{0.001} & --             & \textbf{0.023} & \textbf{0.001} & \ul{0.004}    & --             \\
\rowcolor{gray!10}
AVRG     & \textbf{0.072} & \ul{0.010}    & \textbf{0.007} & --             & 0.026          & 0.042          & 0.374          & 0.263          & \ul{0.356}    & 0.067          & 0.006          & --             & \ul{0.025}    & \ul{0.006}    & \textbf{0.003} & --             \\
\bottomrule
\vspace{.1cm}
\end{tabular}}
    \vspace*{2pt}
\end{table*}

\para{Methodology.} We compared AVRGs against seven state-of-the-art graph generators. Many existing graph generators cannot handle attributed graphs, and many attributed graph generators do not scale to large sizes. The AVRG model can easily scale to extremely large graphs, but we opt to compare against other attributed models necessitating smaller graphs in this evaluation. Specifically, we compared against graph neural network models CELL~\citep{rendsburg2020netgan}, NetGAN~\citep{bojchevski2018netgan}, Graph Autoencoders~\citep{kipf2016variational}, and Variational Autoencoders~\citep{simonovsky2018graphvae}. We also compared with classical, non-neural models such as the Degree Corrected Stochastic Block Model (DC-SBM)~\citep{karrer2011stochastic}, AGM~\citep{pfeiffer2014attributed}, and the base Chung-Lu model~\citep{chung2002average}. We used the best-performing hyperparameter settings for each of the baseline models. 

All experiments were repeated ten times. The mean average of each metric for each model is reported. The 95\% confidence intervals are not shown for clarity but usually lie within 5\% of the mean. The AVRG model used the Leiden clustering algorithm to build the dendrogram, the size-based node selection strategy with $\mu=5$, and the Mixing Matrix rewiring policy for all results in this section.

\subsection{Case Study II: CABAM Simulations} \label{sec:cabam}
The CABAM model simulates scale-free graphs with varying attribute assortativity via a modified, attribute-aware preferential attachment process~\citep{shah2020scale}. We generate graphs with $500$ nodes and degree distribution power-law exponent $\sim$3, and vary the inter-class assortativity ($p_c$) in increments of 0.05. Figure \ref{fig:cabam} shows that for a majority of the attribute assortativity space, AVRG with the Hyperbolic hierarchical clustering (Leiden produces slightly worse results) can produce graphs matching the true values. 
We observe that AVRGs and CELL are the standout performers, both tightly hugging the assortativity curve. Other graph models, including NetGAN (surprisingly), do poorly on this task. These results demonstrate that AVRG is capable of handling graphs across the spectrum of attribute assortativity. 

\subsection{Real-World Graphs}
We evaluated the generation quality of the models by comparing both topological and attribute similarities with the original graphs. 

We choose $\lambda$-distance to compare the topological fidelity instead of comparing degree and PageRank distributions because it is known to be more sensitive to topological variations~\citep{wilson2008study}.
We also computed the absolute difference between degree and attribute assortativity values to quantify similarity in mixing patterns. 
Finally, we counted all \textit{connected} and \textit{colored} 2, 3, and 4-node graphlets (\eg, edges, triangles, squares, kites)~\citep{wernicke2006fanmod}.
We used the inverse of Pearson's correlation coefficient to compare each colored graphlet count vector (the inverse is taken so that lower is better--consistent with our other measures) as a way of measuring higher-order attribute mixing patterns. Because of the enormous number of possible colored graphlets, some of these counts could not be computed for the larger graphs. 

The results on all 12 input graphs are aggregated in~\autoref{tab:topology}. Some models failed due to a combination of memory and model fitting issues. However, we can make the following observations.

CELL and AVRG are the standout performers, both consistently finishing in the top 2. CELL, being the state-of-the-art, performed admirably across all the datasets.  
AVRG does particularly  well on disassortative graphs like Texas, Wisconsin, and Chameleon, in capturing topological and attribute mixing patterns. NetGAN, DC-SBM, AGM, and CL perform similarly well, while the autoencoders fail to reproduce the graphs faithfully.



\section{Conclusions and Future Work}

The present work describes the attributed vertex replacement grammar (AVRG) model inspired by the context-free grammar formalism widely used in compilers and natural language processing. We described how an \update{AVRG} can be extracted from a hierarchical clustering of an attributed graph and then show that the model successfully encodes local topological structures. Starting with an empty graph, if we apply grammar rules stochastically, the model can generate a new graph. We show that the newly generated graphs are similar to the input graph, both topologically and in attribute mixing patterns.

Despite their straightforward design (\ie, lacking supervision and a rich neural architecture), we demonstrate that AVRGs can often surpass state-of-the-art deep graph neural network generators' performance. We further demonstrate that the attributed graph rewriting rules encoded by the grammar can be interpreted to understand the local mixing patterns of the graph and how these substructures are pieced together.

These results open an exciting avenue for further exploration into attributed rewriting rules. One of the principal strengths of AVRGs is their ability to find the same rule repeatedly. For attributed graphs, combining identical grammar rules via rule isomorphism becomes less likely. As the number of unique attributes grows, the model size is likely to grow as well. Combining similar rules using graph kernel methods or other formalisms borrowed from formal language theory may be a particularly compelling avenue for further research. Additional work is also needed to extract actionable knowledge and interpret the simple patterns encoded in the grammar. Such tools could also be used to better understand specific graph datasets such as knowledge graphs, biological networks, and other complex systems containing a rich mixture of attributes and their patterns. We also intend to apply these techniques to solve additional downstream tasks like link prediction, graph summarization, and other critical graph modeling tasks.

\begin{acks}
This work is supported by a grant from the US National Science Foundation (\#1652492).
\end{acks}

\clearpage




\end{document}